\documentclass[aps,twocolumn,prl]{revtex4}
\usepackage{graphicx}
\usepackage{epsfig}
\usepackage{amsmath}
\usepackage{graphics}
\usepackage{latexsym}
\usepackage{xcolor}
\begin{document}

\title{Crystal nuclei in melts: A Monte Carlo simulation of a model for attractive colloids}

\author{Antonia Statt$^{1,2}$, Peter Virnau$^1$ and Kurt Binder$^1$}

\affiliation{$^1$Institut f\"ur Physik, Johannes Gutenberg-Universit\"at Mainz, Staudinger Weg 9, D-55099 Mainz, Germany\\
$^2$ Graduate School of Excellence Materials Science in Mainz, Staudinger Weg 9, 55128 Mainz, Germany}

\begin{abstract} 
As a model for a suspension of hard-sphere like colloidal particles where small nonadsorbing dissolved polymers create a depletion attraction, we introduce an effective colloid-colloid potential closely related to the Asakura-Oosawa model but that does not have any discontinuities. In simulations, this model straightforwardly allows the calculation of the pressure from the Virial formula, and the phase transition in the bulk from the liquid to crystalline solid can be accurately located from a study where a stable coexistence of a crystalline slab with a surrounding liquid phase occurs. For this model, crystalline nuclei surrounded by fluid are studied both by identifying the crystal-fluid interface on the particle level (using suitable bond orientational order parameters to distinguish the phases) and by ``thermodynamic'' means. I.e., the latter method amounts to compute the enhancement of chemical potential and pressure relative to their coexistence values. We show that the chemical potential can be obtained from simulating thick films, where one wall with a rather long range repulsion is present, since near this wall the Widom particle insertion method works, exploiting the fact that the chemical potential in the system is homogeneous. Finally, the surface excess free energy of the nucleus is obtained, for a wide range of nuclei volumes. From this method, it is established that classical nucleation theory works, showing that for the present model the anisotropy of the interface excess free energy of crystals and their resulting nonspherical shape has only a very small effect on the barrier.
\end{abstract}

\maketitle

\section{Introduction}

Understanding how crystalline solids form from fluid phases is a problem that both poses significant intellectual challenges and is a basic input for materials science \cite{1,2,3,4,5,6,7,8,9}. The standard picture implies that nanoscopic nuclei of the solid are formed due to spontaneous thermal fluctuations when the fluid density has a value exceeding the fluid density at fluid-solid coexistence. Due to the cost in surface excess free energy caused by the crystal-fluid interface \cite{4,5}, a free energy barrier in phase space needs to be overcome in every event of this so-called homogeneous nucleation \cite{1,3,4,7} process. In the present paper, we are neither concerned with the further crystal growth processes \cite{2,8,10} that follow after the crystal nuclei have been formed, nor shall we consider the fact that in practice heterogeneous nucleation \cite{8,11,12,13,14,15,16,17,18,19,20,21} may dominate (since the free energy barrier may be significantly reduced for nuclei attached to the walls of the container, or to seed particles \cite{8,13,22,24,25}, etc.).

Due to the large value of the nucleation barrier (which in typical cases is several ten's of the thermal energy $k_BT$ \cite{1,4,5,7}) and the nanoscopic size of these nuclei (typically they contain only a few hundred particles \cite{1,4,5,7}) nucleation events are rare and difficult to observe directly (either in experiment or in simulation). Also the detailed properties of the nuclei that are formed are non-universal and depend upon the particular system under study. Experiments often infer the nucleation rate indirectly from later stages of the process, when the phase transformation from fluid to solid reaches completion \cite{3,4,5,6,7,8,9,10,11,12,13,14}. Of course, this procedure inevitably involves questionable assumptions. Simulations of nucleation kinetics, however, often are feasible only under conditions where the barrier is only of order 10$k_BT$ or less \cite{26,27}. Then it is inevitable that the nuclei are extremely small and strongly fluctuating objects, for which the ``classical'' description in terms of a competition between bulk and surface free energies \cite{1,4,5,7} fails \cite{26,28}.

Many of the difficulties alluded to above are already encountered when one considers nucleation of fluid droplets from a supersaturated vapor \cite{1,4,26}. However, for nucleation of crystals from melts there occur three additional complications (i) the interface tension $(\gamma(\tilde{n}))$ between a crystal surface and the fluid depends on the orientation of the unit vector $\vec{n}$ normal to the surface relative to the axes of the crystal lattice \cite{2,10}. The shape of the nucleus then is not a spherical droplet as in the case of vapor to liquid nucleation, but also exhibits anisotropy. For strong enough anisotropy, facetted nanocrystals are expected \cite{29,30,31,32,33,34,35,36}. In general, the equilibrium crystal shape is nontrivial and can only be obtained from $\gamma(\vec{n})$ via the Wulff construction \cite{29,30,31,32,33,34,35,36}. (ii) Also kinetic processes, such as attachment of particles to the surface of a nanocrystal, may depend on the surface orientation \cite{10}. If the nucleation barrier is not very high, and nucleation plus growth proceed relatively fast, crystal shapes rather different from the equilibrium shape could result \cite{8}. (iii) Often, a fluid at the melting/crystallization transition is rather dense, and enhancing the density further a glass transition is encountered \cite{3}. Then the variation of the kinetic prefactor in the nucleation rate with density becomes an important factor \cite{3,37}. Far from the equilibrium coexistence conditions a separation of a static Boltzmann factor involving the nucleation barrier and a kinetic prefactor may be impossible. Similar problems arise in other systems where the liquid to solid transition is not driven by enhancing the density, as is done in the generic case of colloids described as a hard sphere fluid \cite{9,27,38,39,40}, but by reducing the temperature: in fluids such as silica \cite{3,41,42}, triphenyl phosphite \cite{43,44} and ortho-terphenyl \cite{45} strong supercooling is possible and ultimately a glass transition \cite{3,42} is reached. Nucleation competing with glass formation \cite{3} shall not be considered here, and also the interplay of crystallization with a liquid-liquid unmixing, as it has been proposed for water and various other liquids \cite{44,46} is out of our focus.

The present paper hence is only concerned with the first problem, of predicting the nucleation barrier associated with the formation of crystalline nuclei, but avoiding the assumption that $\gamma(\vec{n})$ actually is isotropic and the nuclei are spherical. A central element of the present paper, however, is, that the nucleation barrier can be obtained directly from a careful analysis of the equilibrium conditions for the nucleus and the surrounding fluid phase \cite{47}. While in the thermodynamic limit this surrounding fluid phase is metastable, we show that in finite systems there actually occurs a stable equilibrium between the small crystalline nucleus and this surrounding fluid phase. We show, using a simple model for a suspension of colloids with short-range attractive interactions, that in this method there is no longer any need to obtain $\gamma(\vec{n})$ and to carry out the Wulff construction. While the precise shape of the crystalline nucleus is not found in this approach, nucleation barriers can be estimated with reasonable precision.

In the next section, we shall describe the model that we use, a variant of the Asakura-Oosawa (AO) model \cite{48,49,50} for colloid-polymer mixtures \cite{51,52}, and summarize what is known on the thermodynamic properties of this model in the bulk. We recall here that the AO model contains the widely studied case of the hard sphere fluid as special case, but has the merit that the control parameters of the attractive part of the potential (its depth and its range, respectively) can easily be tuned (experimentally this could be done by varying the concentration of polymers in the suspension, and their size via choosing different molecular weights \cite{51,52}). Since in this way both the density gap $\rho_m-\rho _f$ (onset of melting of the crystal occurs at a density $\rho = \rho_m$, onset of freezing of the fluid occurs at $\rho=\rho_f$) and the interfacial tension $\gamma(\vec{n})$ can be varied over a wide range, the system qualifies as a model suitable to stringently test the theory.

In section 3, the preparation of the inhomogeneous systems, where a crystalline nucleus coexists with surrounding fluid in a finite simulation box, is described in detail, since the equilibration of such systems exhibiting two-phase coexistence is notoriously difficult \cite{20,28,53,54,55,56,57,58}. We also introduce the local bond orientational order parameters \cite{59,60} that are used to identify (in each system configuration that needs to be further analyzed) where the crystalline nucleus (and its interface separating it from the surrounding fluid) is located. Note that the identification of interfaces between coexisting phases at the single particle level involves important ambiguities \cite{61,62,63,64,65,66}; we feel that previous work where the estimation of the nucleation barrier as function of the nucleus size was attempted, based on such methods, possibly somewhat suffered from this ambiguity.

In Sec.~4 the thermodynamic method utilized in the present paper, where the density $\rho_\ell$, pressure $p_\ell$, and chemical potential $\mu_\ell$ of the liquid surrounding the crystal nucleus are accurately estimated, for a wide range of total particle densities $\rho$ (varying the simulation box volume $V_{box}$ for several choices of the number $N$ of colloidal particles). We explain how from an analysis of this equilibrium we can infer both the volume $V^*$ of the crystalline nucleus, and the associated free energy barrier $\Delta F^*$.

Finally, in Sec. 5 we give a brief summary of our results, and an outlook on further related problems.

\section{The Model and its Bulk Properties}
\subsection{From the Asakura Oosawa (AO) model to the Soft Effective Asakura-Oosawa (Soft EffAO) model}

\begin{figure}
\centering
\includegraphics[width=0.35\textwidth]{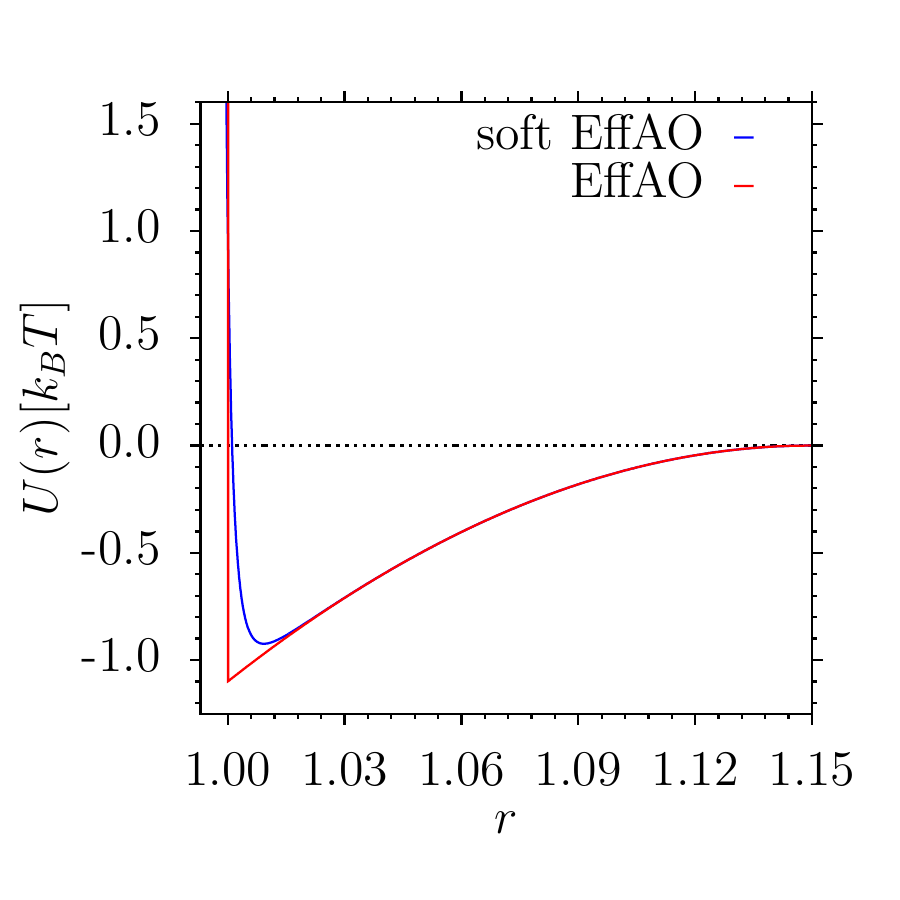}
\caption{\label{fig1} Potential $U(r)/k_BT$ versus distance $r$ between two colloidal particles, for the SoftEffAO model, and compared to $U_{eff}(r)/k_BT$ for the EffAO model. In both cases, $q=0.15$ and $\eta_p^r = 0.1$ is chosen, and $\sigma_c =1$ is used as unit of length.}
\end{figure}

Colloid-polymer mixtures \cite{50,51,52} contain three ingredients: colloidal particles, hard sphere-like in a size range from 10 nm to 1$\mu m$; flexible polymers, which have a high enough molecular weight to form coils in solution with a radius in the range from 10$nm$ to 100$nm$; and a small molecule fluid in which the colloidal particles are suspended, and which must be chosen such that it acts as a good solvent for the polymers or a Theta solvent, in which the polymers take a random walk-like structure \cite{67}. By suitable coating of the colloids with a surfactant layer one can avoid polymer adsorption on the colloids, and screen out the attractive van der Waals interaction between the colloids, avoiding their aggregation.

The AO model provides a simplified description of the statistical thermodynamics of such a system, representing the colloids as hard spheres of diameter $\sigma_c$, and the polymers as soft spheres of diameter $\sigma_p$. While polymer-colloid overlap is also strictly forbidden, polymers can overlap each other with zero energy cost \cite{48,49,50}. As is well known the depletion effect causes an (entropic) effective interaction between the colloids. Note that the molecular solvent in this model is not explicitly regarded at all. If the size ratio $q=\sigma_p/\sigma_c$ is small enough $(q <q^* =2/\sqrt{3}-1 \cong0.1547$ \cite{68}), one can derive an equivalent one-component model (which we refer to as ``Effective AO model'' in the following ) of a colloidal fluid, where the particles interact with the following effective potential:
{\small
\begin{align}\label{eq1}
U_{eff} (r<\sigma_c) &=\infty, \nonumber \\
\beta U_{eff} (\sigma _c <r \leq \sigma_c + \sigma_p) &= \frac \pi 6 \sigma _p^3 z_p (1+q^{-1})^3 \cdot \\ 
& \left(1-\frac{3r/\sigma_c}{2(1+q)}+ \frac {(r/\sigma_c)^3}{2(1+q)^3} \right) ,
\nonumber \\
\textrm{and} \quad  U_{eff} (r \geq \sigma_c+\sigma _p)&=0.
\end{align}}
Here $r$ is the radial distance between two colloidal particles, and $z_p=\exp (\mu_p/k_BT)/\Lambda_p^3$ the polymer fugacity ($\mu _p$ is the polymer chemical potential and $\Lambda _p$ their thermal de Broglie wavelength). In the following we shall denote the model defined by Eq.~\ref{eq1}, where the polymer degrees of freedom were integrated out and hence all information on the distribution of polymers is lost, as Effective Asakura-Oosawa (EffAO) model.

While Eq.~\ref{eq1} is convenient for analytical work (e.g. using density functional theory \cite{69}), it is cumbersome to use in simulations: the singularity of $U_{eff}(r)$ at $r=\sigma _c$ precludes the use of the Virial formula \cite{70} to compute the pressure in the system (alternative methods to compute the pressure in systems containing hard particles exist \cite{71,72} but are difficult to apply and require huge numerical effort; also the use of Molecular Dynamics methods \cite{73,74} would be very inconvenient).

Since real colloidal particles anyway never behave as strictly hard spheres \cite{75,76}, however; there is no need to stick exactly to the potential, Eq.~\ref{eq1}, but one can equally well consider a model where the singular behavior at $\sigma = r_c$ has been replaced by a smooth function. Thus we replace the hard core part of Eq.~\ref{eq1} by the following model, called the Soft Effective Asakura-Oosawa (SoftEffAO) model, while the other parts of Eq.~\ref{eq1} are maintained,
\begin{eqnarray}\label{eq2}
U_{rep}(r)= &4\Big[(\frac{b\sigma_c}{r-\epsilon \sigma_c})^{12} + (\frac {b\sigma _c}{r-\epsilon \sigma_c})^6 -\nonumber\\
&\quad (\frac {b\sigma_c}{\sigma_c +q-\epsilon \sigma_c})^{12}- (\frac {b\sigma_c}{\sigma_c+q-\epsilon \sigma _c})^6\Big]\;.
\end{eqnarray}

The constants b, $\epsilon$ are chosen as $b= 0.01$ and $\epsilon = 0.98857$; of course, the form chosen for Eq.~\ref{eq2} is rather arbitrary, other functions could be used equally well.
Fig.~\ref{fig1} gives a comparison of both potentials, Eq.~\ref{eq1} and Eq.~\ref{eq2}.

\begin{figure}
\centering
\includegraphics[width=0.35\textwidth]{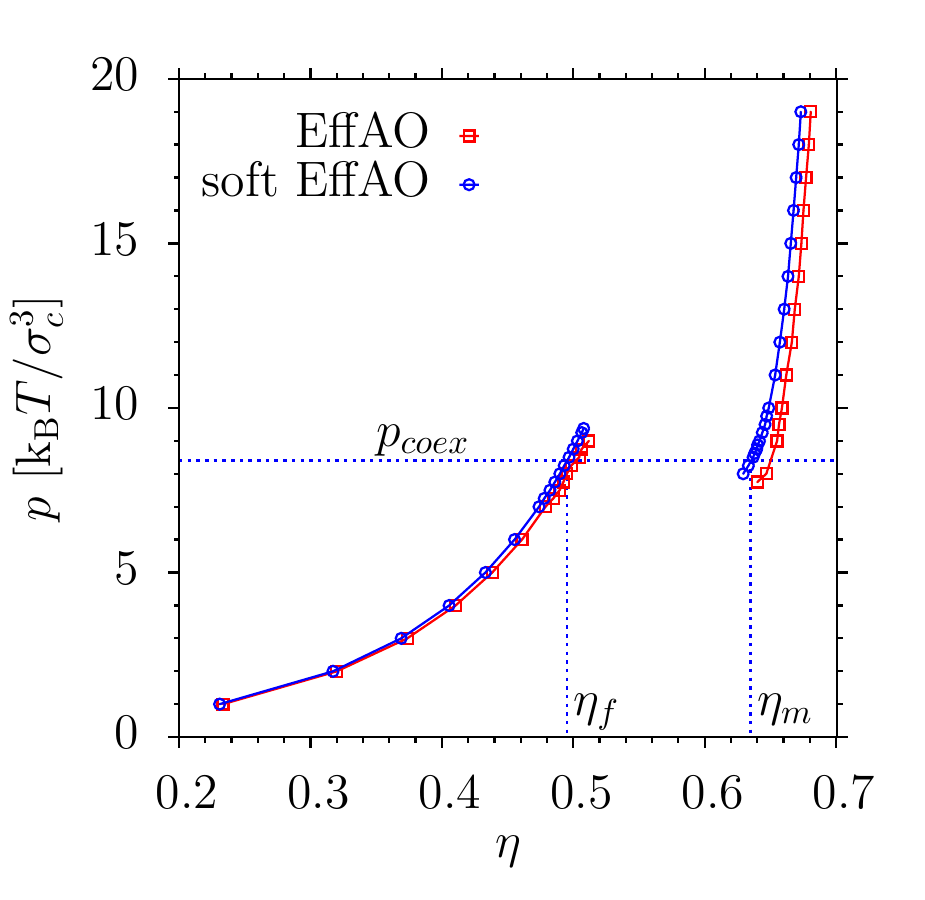}
\caption{\label{fig2} Normalized pressure $p \sigma _c^3/k_BT$ plotted vs. $\eta$ for both the EffAO model (squares) and the soft  EffAO model (circles). Full curves through the data points are guides to the eye only, and statistical errors are smaller than the size of the symbols, and hence not shown. Note that the equation of state contains two disjunct functions; one for the liquid branch $(p_\ell(\eta)$) and one for the crystalline solid ($p_c(\eta)$) (which has a face-centered cubic (fcc) structure), respectively. Note that the data for the liquid branch comprise also metastable states (for $p>p_{coex}$) and the data for the solid branch comprise metastable states as well (for $p<p_{coex}$). The coexistence pressure $p_{coex}$ is found from the interface velocity method (see Fig.~3), and the coexistence packing fractions then result from  $\eta_f = \eta_\ell (p_{coex}), \eta_m = \eta_c(p_{coex}),$ with $\eta_\ell(p), \eta_c(p)$ being the inverse functions of $p_\ell(\eta), p_c(\eta)$. The result for the liquid were obtained using $N = 4000$ colloidal particles and using both the NpT ensemble and the NVT ensemble (see text) to show that finite size effects are negligible.}
\end{figure}

For this SoftEffAO model the application of the Virial formula is straightforwardly possible, and as shown in Fig.~\ref{fig2} , the equation of state of both models is very similar, for the choice $q=0.15$, and $\eta_p^r=0.1$ that will be studied here. Here we have defined the packing fraction $\eta$ of the colloids and the polymer reservoir packing fraction $\eta_p^r$ as $(\rho=N/V_{box})$
\begin{equation}\label{eq3}
\eta =(\pi \sigma _c^3 /6)\rho, \; \eta_p^r=(\pi \sigma _p^3/6) \exp(\mu_p/k_BT).
\end{equation}

Note that $\eta_p^r$ is defined such that it would give the packing fraction of polymers $(\eta_p^r=(\pi\sigma_p^3/6)N_p/V_{box})$ if the simulation box would contain the ideal gas of polymers only.

\subsection{Bulk phase behavior of the SoftEffAO model}

The pressure isotherm shown in Fig.~\ref{fig2}  contains two branches, the liquid branch (at the left) and the solid branch (representing a face-centered cubic (fcc) crystal, at the right). Actually the solid branch was computed in the NpT ensemble, choosing a cubic box with $N = 4000$ particles, arranged at the sites of a fcc lattice as initial condition. Simulations where only the total volume $V=L_xL_yL_z$ is allowed to fluctuate or where the simulation box linear dimensions $L_x,L_y,L_z$ in the three cartesian directions are allowed to fluctuate independently gave (within statistical errors) identical results for the function $\eta = \eta (p)$ that is shown in Fig.~\ref{fig2}.
In our simulations, we have disregarded the alternative hexagonal close-packed (hcp) structure throughout, since it is not expected to be the stable phase~\cite{fcc1,fcc2,fcc3,fcc4}. Of course, it nevertheless would be interesting to study crystal nuclei having the hcp structure surrounded by fluid, applying the technique of the present work, and to compare their nucleation barriers to those of fcc nuclei.
However, this interesting issue must be left to future work.

The liquid branch was obtained from both runs in the $NpT$ ensemble and in the NVT ensemble (sampling then $p$ at given fixed $\eta$ from the virial expression \cite{70,73,74}). Within statistical errors, the results from the NpT and NVT ensembles agree precisely, indicating hence that finite size effects on the isotherms for $N=4000$ are already negligible. We also note that for the ranges of $\eta$ shown we never could observe a spontaneous crystallization of the liquid, nor a spontaneous melting of the solid. This implies that the time span of the simulation runs was too short to observe a decay of a metastable phase via nucleation and growth.

It also is interesting to ask for the effect of choosing different parameters $b$ and $\epsilon$ in the potential, Eq.~\ref{eq2}. One can show that increasing $b$ has the effect of displacing the liquid branch slightly (and almost uniformly) to the left. The effect on the solid branch is much more pronounced, however, it is shifted to the left by much larger extent, and hence the width of the liquid-solid coexistence region decreases \cite{77}.

In thermal equilibrium, the liquid phase and the crystal in Fig.~\ref{fig2}  can coexist at a single pressure only, $p=p_{coex}$, where also the chemical potentials $\mu_\ell(p), \mu_c(p)$ of both phases are equal,
\begin{equation}\label{eq4}
\mu_\ell(p_{coex})=\mu_c(p_{coex})= \mu _{coex}
\end{equation}

\begin{figure}
\centering
\includegraphics[width=0.35\textwidth]{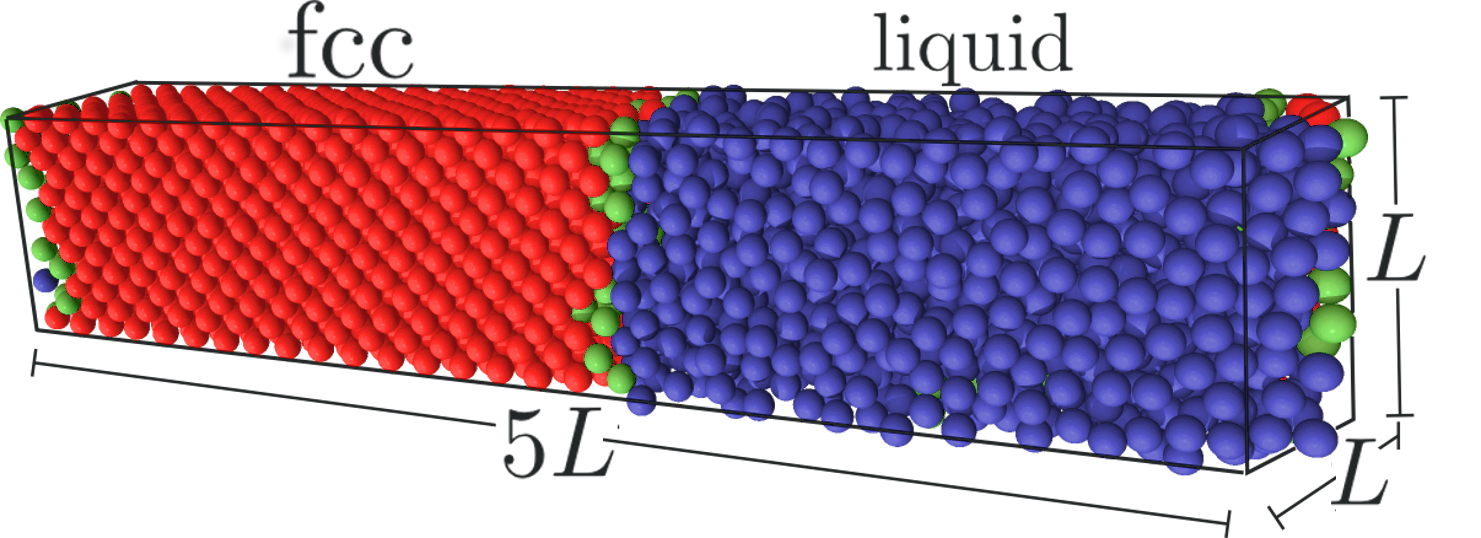}
\caption{\label{fig3} Snapshot picture of a EffAO configuration of a system with linear dimensions $L \times L \times L_z$ with $L_z = 5L$, the crystal slab on the left side and the liquid on the right side. Here $n =6$ and $N=4000$ was chosen, appropriate for $p=8.0$. Particles belonging to the liquid are shown in blue color, particles belonging to the crystal are shown in red color, while particles in the center of the interface are displayed in green color. For the cutoff values for $q_6$ and $q_4$ refer to \cite{color}.}
\end{figure}

\begin{figure}
\centering
\begin{minipage}[t]{0.25\textwidth}
\includegraphics[width=\textwidth]{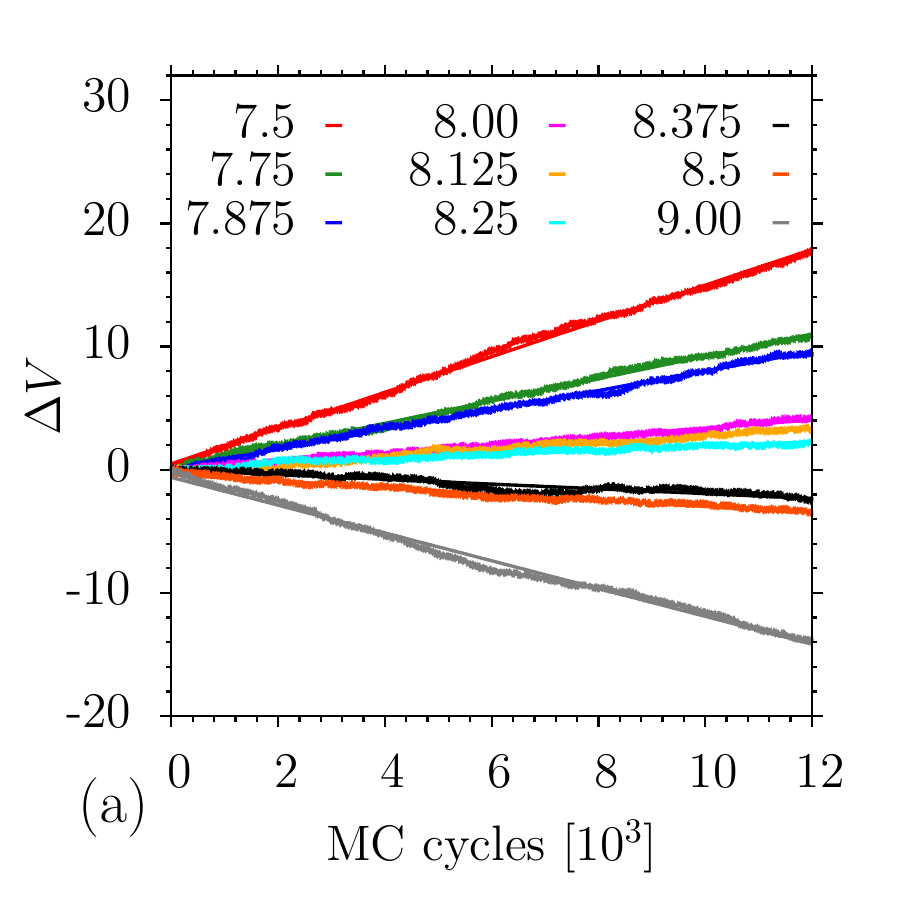}
\end{minipage}%
\begin{minipage}[t]{0.25\textwidth}
\includegraphics[width=\textwidth]{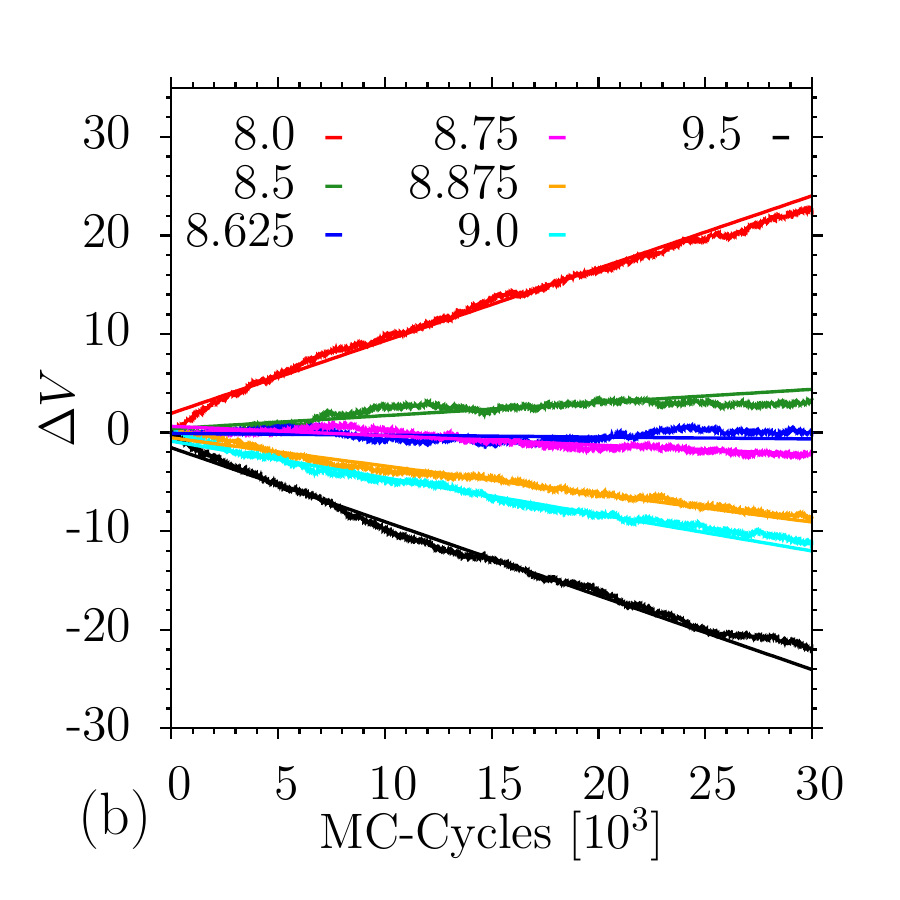}
\end{minipage}
\caption{\label{fig4} Volume change $\Delta V$ as a function of Monte Carlo (MC) cycles for different normalized pressures $p\sigma _c^3/k_BT$, as indicated in the figure, for the effective AO model (a) and for the soft effective AO model (b). The system contains $N=9250$ colloids in each case, corresponding to $n=8$ lattice planes in x and y direction. The solid straight lines are fits, from which $\langle \Delta V/dt\rangle$ is determined. Each curve is averaged over 100 independent runs.}
\end{figure}

\begin{figure}
\centering
\begin{minipage}[t]{0.25\textwidth}
\includegraphics[width=\textwidth]{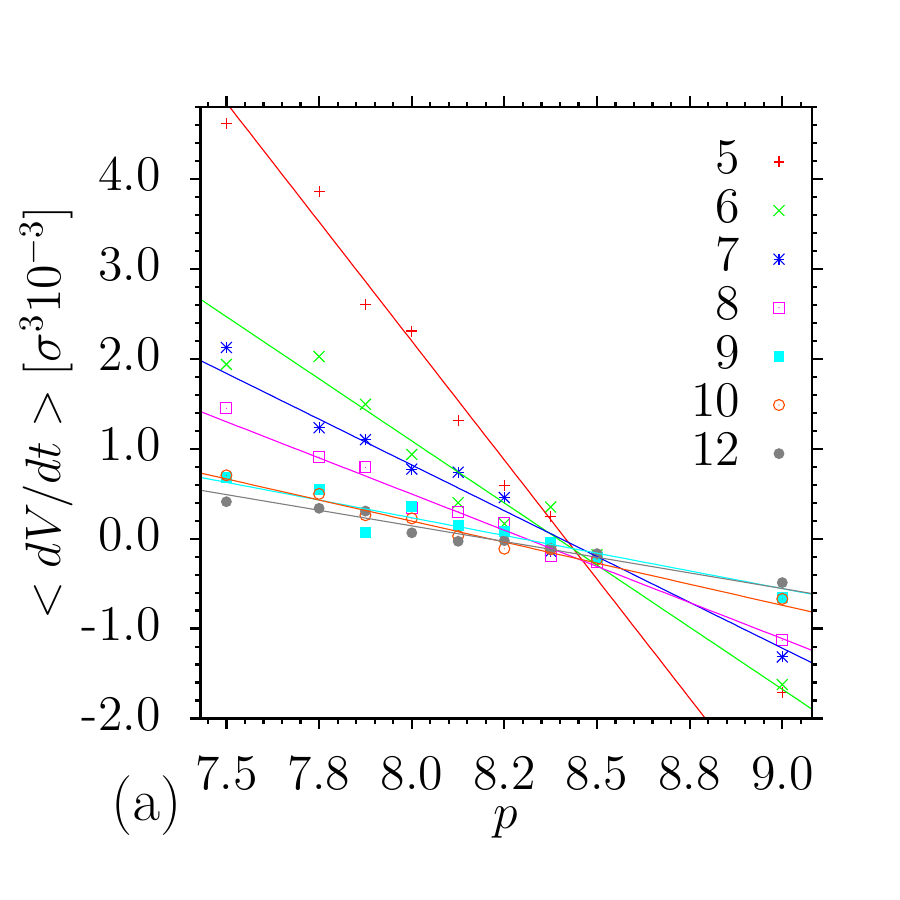}
\end{minipage}%
\begin{minipage}[t]{0.25\textwidth}
\includegraphics[width=\textwidth]{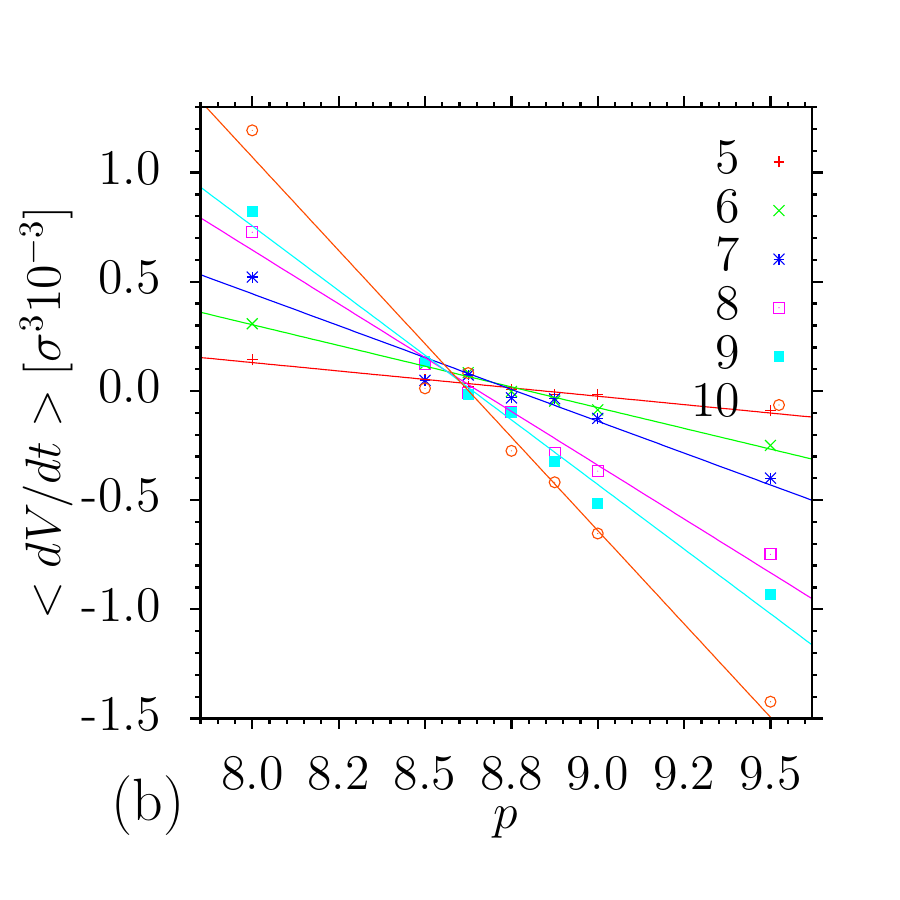}
\end{minipage}
\includegraphics[width=0.25\textwidth]{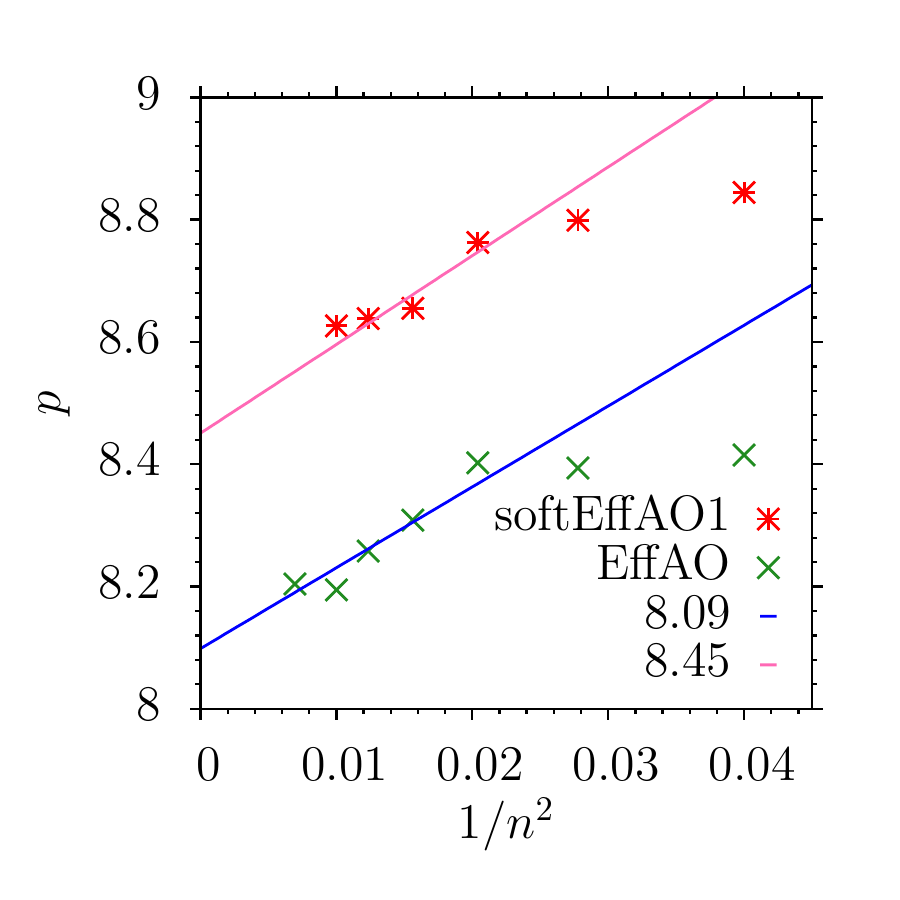}
\caption{\label{fig5} (a) Velocities of volume changes $\langle \Delta V /dt\rangle$, as determined in Fig.~4, plotted vs. $p$, for different choices of $n$, as indicated. Results for the effective AO model and for the soft effective AO model (b) are shown. Straight line fits of the data are shown, from which the transition point $(\langle \Delta V /dt\rangle =0$, highlighted by a horizontal straight line) $p=p_{coex}$ is estimated. (c) Pressure $p$ where $\langle \Delta V /dt\rangle =0$ plotted vs. $1/(n^2)$, where $n$ is the number of lattice planes in $x$ and $y$ direction in the box. Disregarding the smallest size, the remaining data are fitted empirically to straight lines, yielding the result $p_{coex}=8.09 \pm 0.06$ for the EffAO model and $p_{coex} = 8.45 \pm 0.04$ for the soft EffAO model.}
\end{figure}

But from Fig.~\ref{fig2}  we see that there occurs an extended region of hysteresis, where either the liquid $(p> p_{coex})$ or the crystal $(p <p_{coex})$ is metastable in our simulation. Hence the accurate estimation of $p_{coex}$ is a nontrivial problem. We follow the procedure of Zykova-Timan et al. \cite{65} by locating $p_{coex}$ from studies of ``slab configurations'' (Fig.~\ref{fig3}) of liquid-solid coexistence in the NpT ensemble. We estimate the interface velocity from the time derivative of the volume $\langle dV/dt\rangle$ as a function of pressure $p$, identifying $p_{coex}$ from the pressure for which $\langle dV/dt\rangle$ changes sign (Figs.~\ref{fig4},\ref{fig5}). The initial system at each pressure is chosen as rectangular parallelepiped, $V=L\times L\times L_z$, choosing $L_z \approx 5L$, and a crystal slab of extension $L_c(\approx L_z/2)$ in the z-direction, with two (001) interfaces (which for the length of the runs performed are still non-interacting, even if the crystal gradually melts (and therefore the volume change $\Delta V$ increases with time, Fig.~\ref{fig4}). At each $p$, the linear dimension $L$ is chosen as $L=nd(p)$, $n$ being an integer $(5\leq n \leq 12)$ and $d(p)$ the equilibrium lattice constant at the pressure $p$ that we wish to test; $d(p)$ can be estimated from the crystal branch in the $p$ vs. $\eta$ isotherm (Fig.~\ref{fig3}). Also $L_c$ was chosen commensurate with an integer number of crystal planes. Such a choice of linear dimensions commensurate with the crystal structure is essential, since we use periodic boundary conditions throughout, and one must avoid strain deformations of the crystalline slab. The remaining volume $L \times L \times (L_z-L_c)$ contains liquid with a density chosen according to the liquid branch on the isotherm $\eta(p)$ for the pressure that is tested (the precise value of $L_z$ for the initial configuration is chosen such that the particle number $N$ is an integer, of course). To prepare the initial configuration for the NpT simulation of this slab configuration we actually carry out first a run in the NVT ensemble with V fixed through the initial values of $L$ and $L_z$, to achieve that the liquid configurations in the crystal-liquid interfacial region also equilibrates (only particles in the liquid region take part in this initial equilibration, carrying out about 10$^{10}$ Monte Carlo moves, i.e. displacements of single particles).

In the simulations in the $NpT$ ensemble then all particles are considered for displacement moves, and in the volume moves all three linear dimensions $L_x,L_y$ and $L_z$ are allowed to fluctuate independently (unlike \cite{65} where $L_x=L_y=L=nd(p)$ was held fixed, so only $L_z$ could fluctuate), but keeping $n$ fixed. This procedure required an order of magnitude more statistical effort than used in \cite{65}. The length of the runs was $3 \times 10^4$ MC cycles (one cycle consisted of $N$ particle displacement moves and one volume move), and in each case 100 independent runs were made.

Fig.~\ref{fig4} shows that $\Delta V$ increases with time for low pressures (as expected, due to the progressing melting of the crystalline slab the volume fraction of the liquid that requires more volume increases) and for high pressures the opposite trend is observed. Clearly, these data still are affected by statistical fluctuations, and there is also significant scatter in the variation of the interface velocities with pressure (Fig.~\ref{fig5}c). These difficulties are a serious limitation for the accuracy with which $p_{coex}$ can be determined. Taking also the trend with increasing number $n$ of planes (and hence increasing linear dimension $L$) into account (Fig.~\ref{fig5}b), we conclude
\begin{eqnarray}\label{eq5}
p_{coex} = 8.09 \pm 0.06 \; \textrm{(Eff\;AO\; model)},\nonumber \\
p_{coex} = 8.45 \pm 0.04 \; \textrm{(soft\;Eff\; AO\; model)}.
\end{eqnarray}

In our preliminary publication \cite{47}, slightly different values were quoted, $p_{coex} = 8.06  \pm  0.06$ ( EffAO model) and
 $p_{coex} = 8.44 \pm  0.04$  (soft Eff AO) relying on the empirical observation that an extrapolation against $1/\sqrt{N}$ seemed to
work best \cite{47}. Note, however, that there is no theoretical justification for an extrapolation against $1/\sqrt{N}$, one might
rather expect an extrapolation against the inverse interface area should be used, as done in Fig. 5c. As a caveat, we
note that if we would have chosen only the tree smallest choices of $n$, one would have erroneously concluded that
finite size effects on pcoex are very small, resulting in estimates significantly off. 
 As we shall see later, this error (which is less than 1\% ) is the most critical limitation to the accuracy that we can reach for the prediction of nucleation barriers; however, it is by no means straightforward to obtain $p_{coex}$ with significantly smaller errors. Knowing the coexistence pressures the corresponding packing fractions of the coexisting phase are readily found from Fig.~\ref{fig2}. For the soft effective AO model, this implies
\begin{equation}\label{eq6}
\eta_f = 0.495 \; (1) \quad , \quad \eta_m = 0.636  \, (1)
\end{equation}

\subsection{Computing the chemical potential for dense fluids}

The chemical potential $\mu$ of the colloids is a basic quantity for the study of phase coexistence, since (unlike the pressure tensor) $\mu$ is strictly spatially constant in a system in thermal equilibrium, even if the system is inhomogeneous due to the presence of interfaces, external walls, etc. Thus when we consider the coexistence of a crystalline nucleus of finite size with surrounding liquid in a simulation box, we have
\begin{equation}\label{eq7}
\mu = \mu _\ell (p_\ell) = \mu _c (p_c) > \mu_{coex}
\end{equation}

instead of Eq.~\ref{eq4}, which applies in the thermodynamic limit. We expect $p_c >p_\ell>p_{coex}$, due to the Laplace pressure effect. Quantification of these differences $\mu - \mu_{coex}$, $p_\ell -p_{coex}, \; p_c - p_{coex}$ as function of nucleus size is one of the main aims of the present paper, which will allow us also to estimate the nucleation barrier, without identification of the crystal-liquid interface on the particle level (Sec.~4).Therefore precise estimation of the chemical potential is a matter of concern.

Now it is well known that the standard Widom particle insertion method \cite{78} becomes inapplicable at liquid densities near the melting transitions, since the acceptance probability for virtual particle insertions becomes extremely small. It was early recognized, however, that in principle one could circumvent this problem studying a system with a soft repulsive wall, since near this wall the density is reduced and hence the acceptance probability is higher \cite{79}. Since the chemical potential near the wall must be the same as far away from the wall (where the liquid density may be so high that particle insertions fail), we know the chemical potential also in the latter region.

Of course, the accuracy of this approach is a nontrivial problem; the spatial extent of the region where the chemical potential can be ``measured'' covers only a small fraction of the system when the system is large. When the system is rather small, as used in \cite{79}, the density shows pronounced oscillations throughout the simulation box (``layering''), and then it is not at all clear to which bulk conditions the considered system would correspond. As a matter of fact, to our knowledge the simple idea of Ref.~79 has never been convincingly validated with respect to its practical usefulness yet.

\begin{figure}
\centering
\begin{minipage}[t]{0.25\textwidth}
\includegraphics[width=\textwidth]{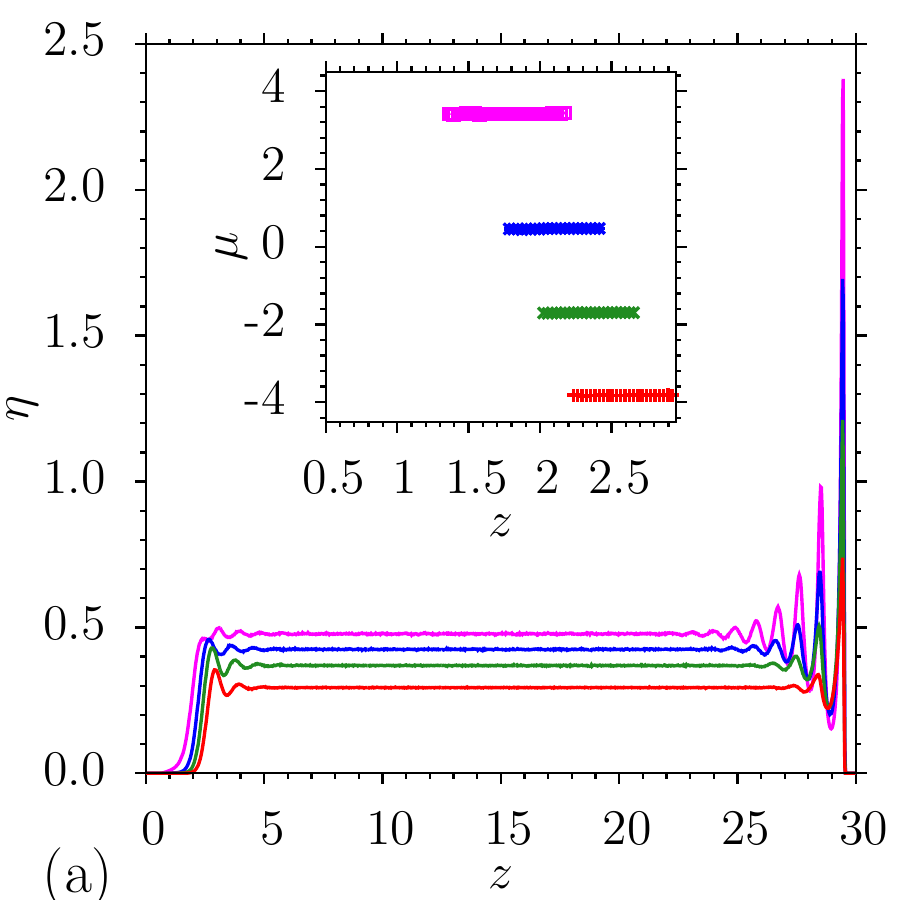}
\end{minipage}%
\begin{minipage}[t]{0.25\textwidth}
\includegraphics[width=\textwidth]{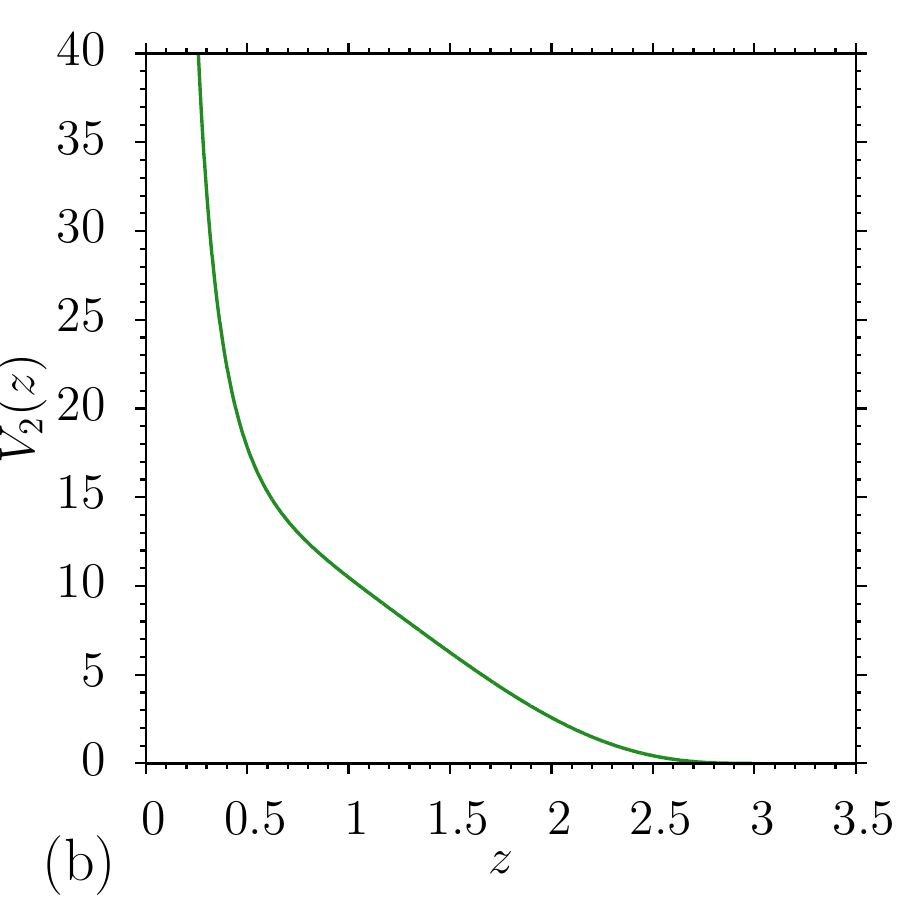}
\end{minipage}
\caption{\label{fig6} 
(a) Illustration of the method to compute the chemical potential of a dense colloidal fluid, using a $L \times L \times L_z$ geometry with two walls at $z=0$ and at $z=L_z$, and periodic boundary conditions in x and y directions, for the case $L=7$, $L_z=30$ and four choices of particle number ($N=750, 950, 1100 $ and 1250, respectively, at $z=15$ from bottom to top respectively). At z = 0 a soft repulsive potential acts (Eq.~\ref{eq3}) while at $z=L_z$ a hard wall acts (which causes significant layering). The insert shows the regions where in each case the chemical potential $\mu$ is computed (and demonstrates that $\mu$ is independent of $z$ in this region in each case, as it should be). The bulk liquid packing fraction corresponding to $\mu$ can be read off from an average over the region where the packing fraction profile $\eta(z)$ is constant, i.e. for $10 \leq z \leq 20$ in this example.
(b) Soft wall potential $V_2(z) $ given in Eq.\ref{eq9} as function of distance  $z$ in colloid diameter to the wall. }
\end{figure}

Fig.~\ref{fig6}  now compellingly shows that the method indeed works well, for our model. At the strongly repulsive wall at the right side in Fig.~\ref{fig6} , for $z=L_z$, we use a Weeks-Chandler-Anderson-type potential
\begin{align}\label{eq8}
V_1 (z) = 4 \epsilon _{w1} [\left(\frac {\sigma_{w1}}{z'}\right)^{12} -\left(\frac {\sigma _{w1}}{z'}\right)^{6} + \frac 1 4 ] ,\\
 z'=L_z-z, \; 0 \leq z' \leq \sigma_{w1}2^{1/6},\nonumber
\end{align}
with $\epsilon_{w1}=1$ and $\sigma_{w1}=\sigma_c/2$, and $V_1(z) =0$ for $z'\geq \sigma_{w1}2^{1/6}$. At the weakly repulsive wall at the left side in Fig.~6, we choose a potential of the form
{\small
\begin{align}\label{eq9}
&V_2 (z) =\epsilon_{w2}\Big(\Big[
\left(\frac{\sigma_{w2} }{ z-\epsilon'_{w2}}\right)^2-\frac{\sigma_{w2} }{ z-\epsilon'_{w2}}+\frac 1 4 
\Big] +B(z) \Big)S(z) \\
&\text{where } B(z)=16\Big[1 - 10\left(\frac{z-\epsilon_{w2}}{z_c-\epsilon_{w2}}\right)^3 + 15\left(\frac{z-\epsilon_{w2}}{z_c-\epsilon_{w2}}\right)^4  \nonumber \\ 
&\qquad \; \qquad \qquad \qquad - 6\left(\frac{z-\epsilon_{w2}}{z_c-\epsilon_{w2}}\right)^5\Big] \\
&\text{and } 
S(z)=\frac{(z - z_c)^4}{(h+(z - z_c)^4} \nonumber
\end{align}}
with $\epsilon_{w2}= 0.8,$ $\epsilon'_{w2}= 0.1,$ $\sigma_{w2} = 1.5,$ $h=10^{-8}$ and $V_2(z) = 0$ for $z >z_c=2.2 \sigma_{w2}$. Note that neither $V_1(z)$ nor $V_2(z)$ exhibit a discontinuity at their cutoff values, also the force is continuous there. Due to the large range of $\sigma _{w2}$ and the rather smooth variation of $V_1(z)$ the layering of the packing fraction $\eta (z)$ quickly reaches a horizontal plateau. Due to the depletion of particles near the soft wall (and the excess density that occurs near the hard wall, where a pronounced layering is observed \cite{70}) the packing fraction $\eta_b$ that is observed in the flat region of $V_2(z)$ differs somewhat from the average packing fraction $\bar{\eta}$ in the simulation box. Note that
\begin{equation}\label{eq10}
\bar{\eta} = L_z^{-1} \int \limits _0^{L_z} \eta (z) dz = N/(L^2L_z)
\end{equation}
In the example shown in Fig.~\ref{fig6} , we have chosen $\bar{\eta} \approx 0.2671, \; 0.3384, \; 0.3918$ and $0.4452$ (from bottom to top), while the corresponding estimates for $\eta_b$ were $\eta_b=0.2933, 0.3690, 0.4241$ and $0.4777$, respectively. The chemical potential $\mu$ (shown in the insert) needs to be associated with $\eta_b$, of course, since $\bar{\eta}$ differs from $\eta_b$ by wall excess corrections,
\begin{equation}\label{eq11}
\bar{\eta} = \eta_b + \frac {1}{L_z} \eta_{w1} \; + \frac {1}{L_z} \eta_{w2}\;,
\end{equation}
with
\begin{equation}\label{eq12}
\eta_{w1}= \int \limits ^{L_z/2}_0 (\eta (z)-\eta_b) dz\quad , \quad \eta_{w2} = \int \limits _{L_z/2}^{L_z} (\eta (z)-\eta_b) dz.
\end{equation}

The data given would allow precise estimation of these excess densities of the walls, $\eta_{w1}$ and $\eta_{w2}$, as well, but this is out of the scope of the present paper. We rather emphasize that near the left wall, where $\eta(z)$ gradually rises to $\eta_b$, we can use a significantly broad slab (of a width $\Delta z$ of the order $\Delta z = 1$) where $\eta(z)$ is small enough that the Widom particle insertion method allows an accurate sampling of the chemical potential. Although $\eta(z)$ is rapidly varying in these intervals of $z$ where the sampling of $\mu$ is done ,we do find that $\mu$ is nicely constant in each case, the statistical fluctuations being smaller than the thickness of the horizontal straight lines shown in the insert.Although this method relies only on a small fraction $(\Delta z/L_z)$ of the volume that is simulated, nevertheless a satisfying accuracy of the function $\mu(\eta_b)$ in the fluid phase is reached.
One could have chosen a soft wall potential at both walls, and thus improve the statistics for the estimation of the chemical potential somewhat. Such a choice would in particular
be necessary in cases where complete wetting of the fluid at melting conditions at the hard walls occur, since then there would be an extended range of strong layering. In our case, the layering extends over a few particle diameters only, and thus does not matter.

\begin{figure}
\centering
\begin{minipage}[t]{0.25\textwidth}
\includegraphics[width=\textwidth]{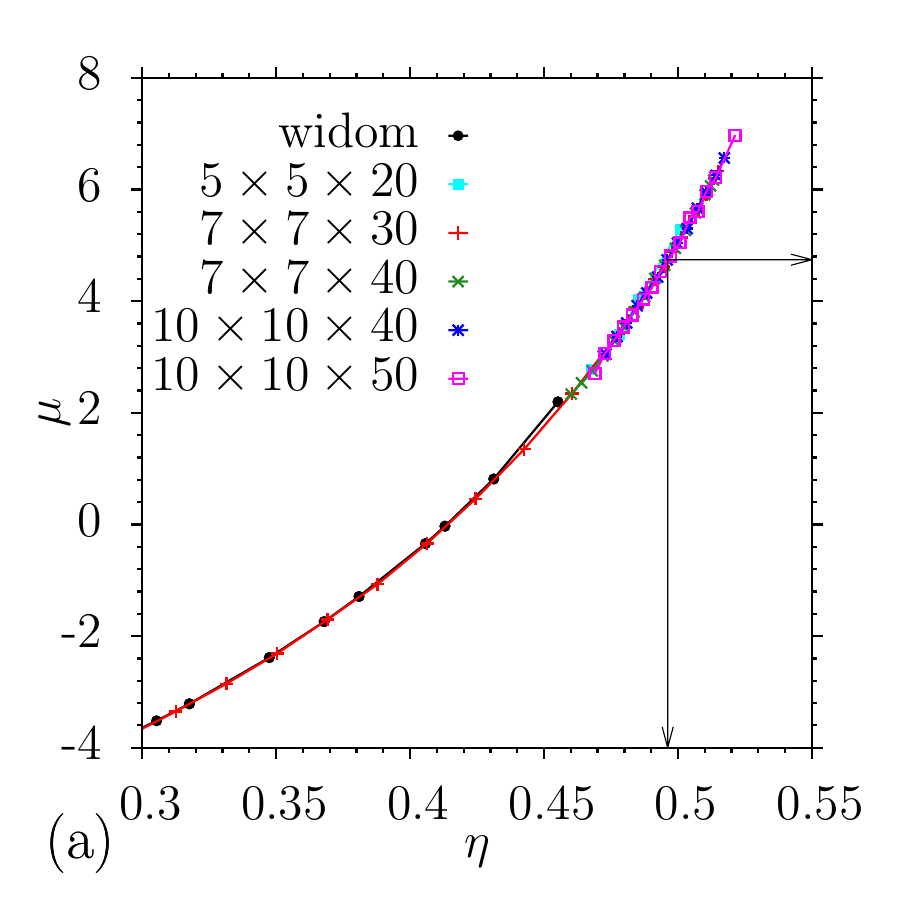}
\end{minipage}%
\begin{minipage}[t]{0.25\textwidth}
\includegraphics[width=\textwidth]{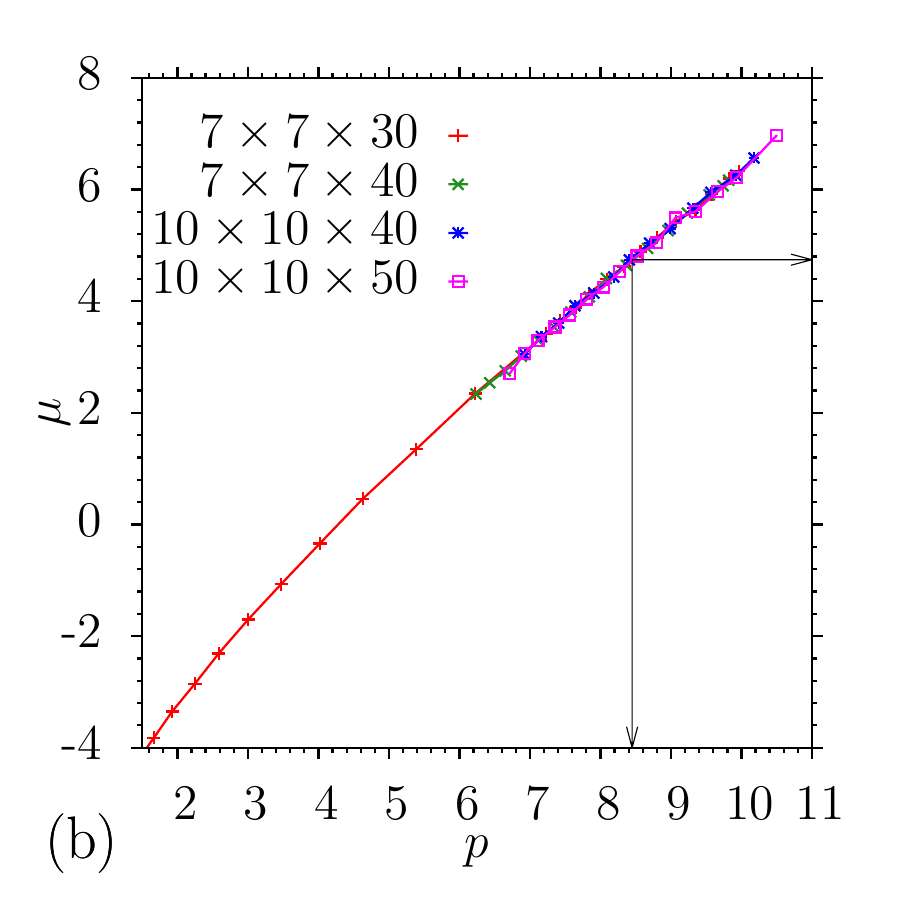}
\end{minipage}
\caption{\label{fig7} (a) Chemical potential $\mu$ (in units of $k_BT$) plotted vs. packing fraction $\eta$, for different choices of $L$ and $L_z$, as indicated. The data labeled as ``Widom'' (available for not so large packing fractions) are obtained by the standard particle insertion method for homogeneous bulk systems. Arrows on abscissa and ordinate indicate $\eta_f$ and $\mu_{coex}$, respectively. (b) Chemical potential $\mu$ plotted vs. pressure $p$, using the same data as in part (a). Arrows at the abscissa and ordinate indicate $p_{coex}$ and $\mu_{coex}$, respectively.}
\end{figure}
Fig.~\ref{fig7}  presents then a plot of $\mu$ vs. $\eta$ over a wide range $(0.3 < \eta < 0.52$) of the fluid branch of the isotherm, including hence also its metastable part. For not too large values of $\eta (\eta <0.46)$ also estimates recorded for bulk $L \times L \times L$ systems (with no walls and periodic boundary conditions throughout) are included, applying the standard particle insertion method. These estimates are in excellent agreement with the results found from the method described above, only for the last point $(\eta \approx 0.455$) the result of this direct method is slightly off, due to a too small acceptance rate of the particle insertions. We also include many different choices of both $L$ and $L_z$ in Fig.~\ref{fig7}, and demonstrate that our results are not hampered by finite size effects. Since we know the transition point $\eta = \eta_f=0.495$ (\ref{eq1}) from Figs.~(\ref{fig2},\ref{fig5}b), we can read off $\mu_{coex} = 4.74$ (\ref{eq4}) as well (Fig.\ref{fig7}a,b). Note that Fig.~\ref{fig7}b shows that the relation $\mu_\ell (p)$ versus $p$ for the liquid branch near $p_{coex}$ is very accurately represented by a straight line, hence validating a simple Taylor expansion

\begin{equation}\label{eq13}
\mu_\ell (p_\ell) \approx \mu_{coex} + \frac \pi 6 \frac {1}{\eta_f} (p_\ell - p _{coex})
\end{equation}

Here we have given the pressure an index $\ell$ to emphasize that Fig.~\ref{fig7}b and Eq.~\ref{eq13} describe the pressure of the liquid phase and not the crystal (when we consider a crystal nucleus surrounded by the liquid phase, the pressure $p_c$ in the crystal exceeds $p_\ell$).

\section{Homogeneous systems versus Two-Phase Coexistence}
\subsection{Local Identification of Phases in Terms of Bond Orientational Order Parameters}

When we want to analyze the phase coexistence between a crystalline nucleus and surrounding liquid phase (the associated chemical potential then can be inferred from Fig.~\ref{fig7}b) as a function of the nucleus volume, it is mandatory to identify where inside the simulation box the crystalline nucleus actually is located, for each configuration of the colloidal particles that is analyzed. Due to statistical fluctuations fluid particles
 in the interfacial region may get attached to the crystal, as well as the inverse process. While in equilibrium there should be just fluctuations of the size and shape of the crystalline nucleus around their proper average values, these fluctuations will also lead to random displacements of the center of mass of the crystalline nucleus. In a long simulation run, we hence should have a diffusive motion of the crystalline nucleus in our system.

Of course, reliable ``measurements'' of $p_\ell$ can only be taken in such subboxes of the system which are not affected by the presence of the nucleus; i.e. remote also from the region of the crystal-liquid interface in the analyzed configuration of the system. Note that also the orientations of the lattice planes at the crystal nucleus can be arbitrary, they should not be biased by the axes of the simulation box, if the simulated system is large enough.

\begin{figure}
\centering
\begin{minipage}[t]{0.25\textwidth}
\includegraphics[width=\textwidth]{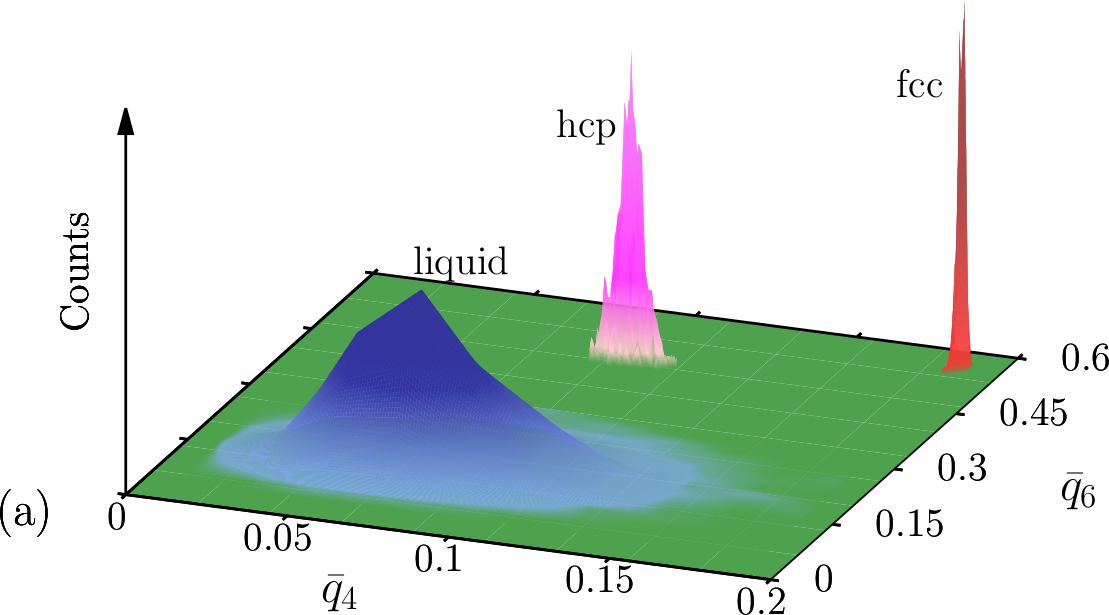}
\end{minipage}%
\begin{minipage}[t]{0.25\textwidth}
\includegraphics[width=\textwidth]{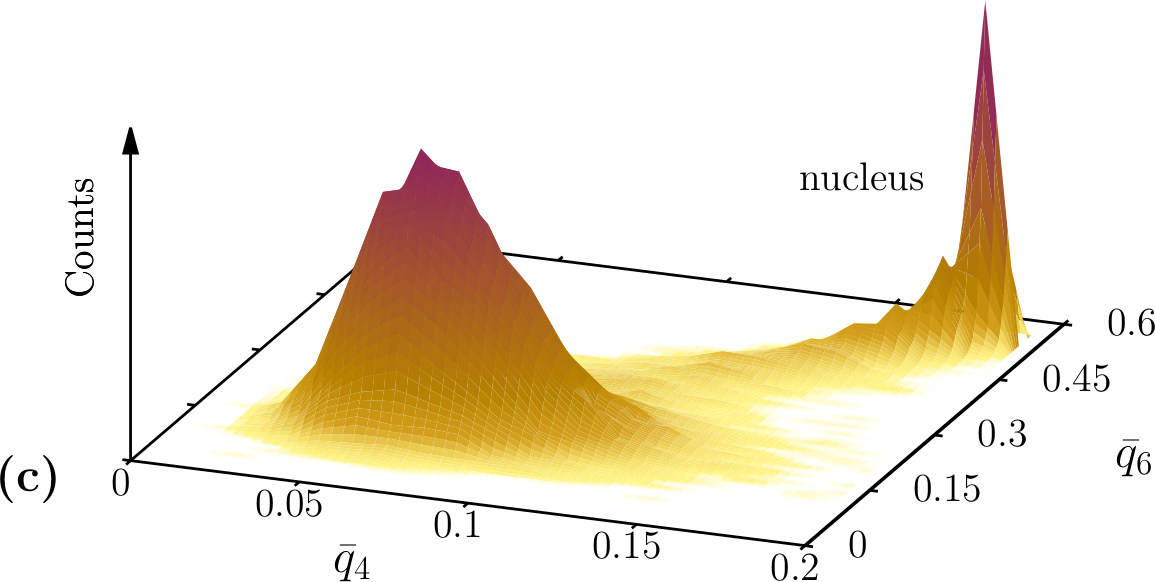}
\end{minipage}
\includegraphics[width=0.25\textwidth]{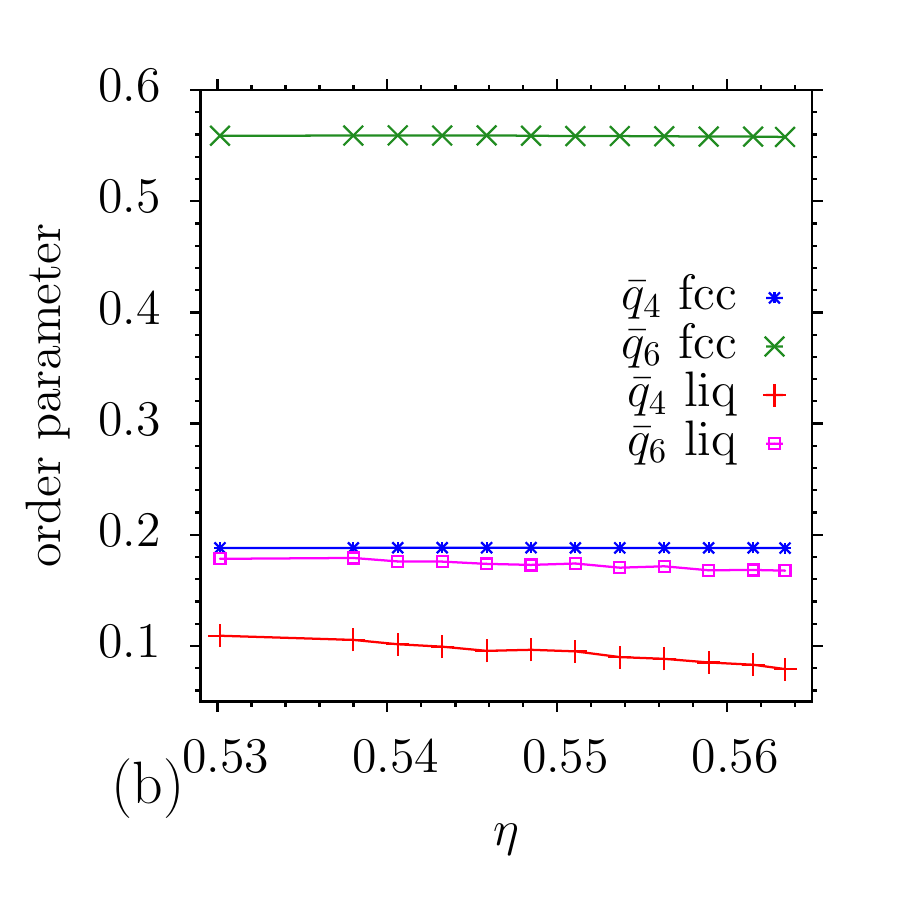}
\caption{\label{fig8} (a) Distribution $P(\tilde{q}_4,\tilde{q}_6)$ in the liquid ( at $\eta = 0.485$ ), hcp and fcc phases (at $\eta= 0.65$) (b) Variation of $\langle \tilde{q}_4 \rangle$ and  and $\langle \tilde{q}_6 \rangle$ with $\eta$ in the liquid and fcc phases, for the range from $\eta  = 0.53$ to $\eta = 0.565$. (c) Distribution recorded for the coexistence of a crystal nucleus surrounded by fluid. This case refers to an average packing fraction $\eta = 0.528$ in the simulation box and $\eta_\ell= 0.512$ in its liquid part.}
\end{figure}
The identification of whether particles are part of the liquid phase or part of the crystalline nucleus is based on the well-known Steinhardt bond orientational order parameters \cite{59,60}. They are defined in terms of spherical harmonics $Y_{\ell m}(\Omega)$
as
\begin{equation}\label{eq14}
q_{\ell m}(i) = \frac {1} {N(i)} \sum \limits ^{N(i)} _{j=1} Y_{\ell m}(\theta (\vec{r}_{ij}), \phi (\vec{r}_{ij}))
\end{equation}
where $\vec{r}_{ij}$ is the distance vector from the considered colloidal particle, labeled by index i, to one of its $N(i)$ neighbors within a cutoff radius (defining what the ``nearest neighbors'' are). The polar angles $\theta (\vec{r}_{ij}) $ and $\phi(\vec{r}_{ij})$ of this distance vector are chosen with respect to an arbitrary but fixed reference frame. We use a cutoff radius of 1.3 $\sigma$ here, determined by the first minimum of the radial distribution function $g(r)$ measured in a bulk fcc crystal. It is advantageous \cite{60} to average the $q_{\ell m} (i)$ over a particle and all its nearest neighbors
\begin{equation}\label{eq15}
\tilde{q}_{\ell m}  (i) = \frac {1} {N(i) +1} \sum \limits _{k _i=0}^{N(i)} q_{\ell m} (k_i),
\end{equation}
where $k_i=0$ means that one takes $q_{\ell m}(i)$ from Eq.~\ref{eq14}. The order parameter that finally is used is then defined as
\begin{equation}\label{eq16}
\tilde{q}_\ell(i) = \left(\frac{4\pi}{(2\ell +1)} \sum \limits _{m=-\ell}^{+\ell} |\tilde{q}_{\ell m} (i)|^2 \right)^{1/2}
\end{equation}
In order to distinguish fcc crystals from liquids it is useful to consider the resulting distribution of $\tilde{q} _4$ and $\tilde{q}_6$ (Fig.~\ref{fig8}a). One sees that in the pure phases the distribution are well separated from each other. Since it is conceivable that in simulations of phase coexistence also hexagonal closed packed (hcp) regions compete with fcc regions, also data for $P(\tilde{q}_4,\tilde{q}_6)$ for a hcp crystal are included: However, in our study we have not encountered any evidence that this structure plays a role for our model. It is also gratifying to note that the peak position of $\tilde{q}_6$ and $\tilde{q}_4$ are basically independent of $\eta$, when one studies fcc crystals in the relevant region of packing fractions (Fig.~\ref{fig8}b).

The situation is somewhat different when one considers phase coexistence between a crystalline nucleus and surrounding fluid (Fig.~\ref{fig8}c): Now one finds a ``ridge'' connecting the sharp peak (representing the crystalline nucleus) with the broad peak representing the liquid. Clearly there is an ambiguity whether particles yielding a ``signal'' in this intermediate region are to be ``counted'' as part of the liquid or as part of the crystal. We shall return to this problem in the next section.

\subsection{Preparation and local Characterization of Systems exhibiting Phase Coexistence Between a Crystalline Nucleus and Surrounding Fluid}

We study systems of cubic shape ($\L \times L \times L$ boxes with periodic boundary conditions) at constant volume and a particle number $N$ chosen such that the packing fraction $\eta$ is inside the two-phase coexistence region, namely $0.525 \leq \eta \leq 0.530$, for a total number of colloids $N = 6000$, 8000 and 10000 and start with a situation where a compact crystal nucleus (of about the right volume $V_c$ and packing fraction $\eta_m$, so that the remaining particles are compatible with a surrounding liquid of the expected density) is initially already present. In the quoted parameter range, one finds that after a (very long) equilibration stage the system settles down at the stable equilibrium situation, where the crystal has somewhat grown or shrunk in size and adjusted its shape, such that the proper equilibrium is established.
\begin{figure}
\centering
\includegraphics[width=0.4\textwidth]{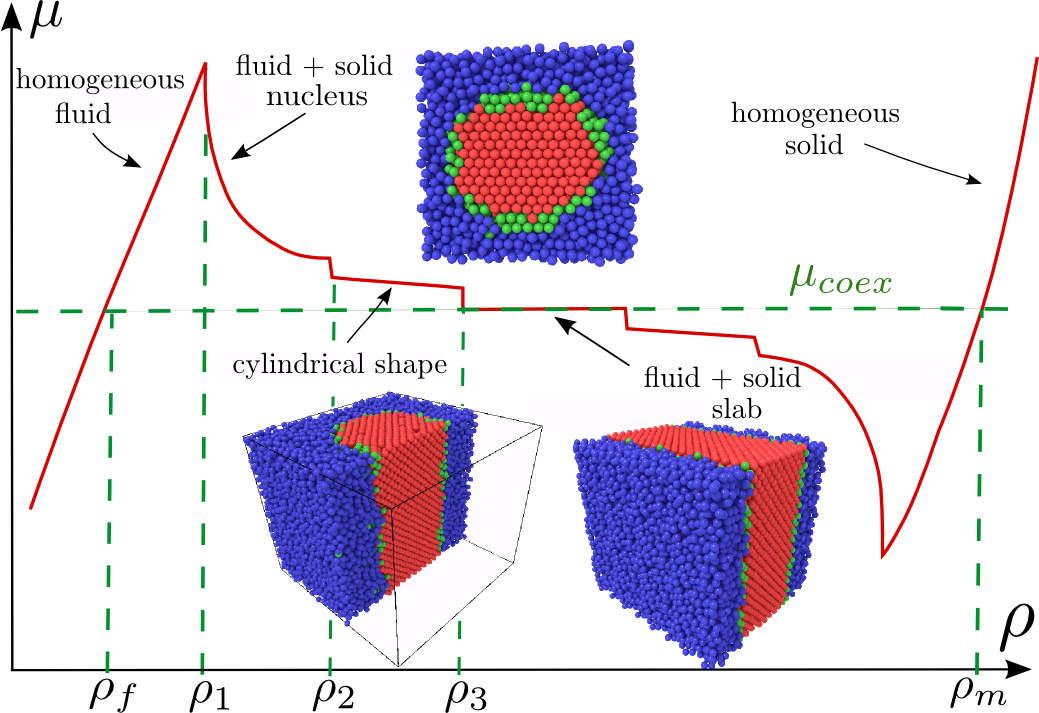}
\caption{\label{fig9}
Schematic plot of the chemical potential $\mu$ versus density $\rho$ for a system undergoing a liquid-solid transition in a finite  box volume $V_{box}$with periodic boundary conditions. Due to interfacial effects, non-negligible in finite systems, the isotherm deviates from $\mu=\mu_{coex}$ in the two-phase coexistence region, $\rho_f<\rho <\rho_m$. The sharp kinks of the curve are in reality rounded due to fluctuations as long as $V_{box}$ is finite (for $V_{box} \rightarrow \infty$, however, all special features of this isotherm disappear and only the horizontal straight line $\mu = \mu_{coex} $ remains). At the density $\rho_1$, the system undergoes  a transition from homogeneous fluid to a situation of two-phase coexistence (crystalline nucleus (red particles) surrounded by fluid (blue particles), see \cite{color}). Particles in the center of the interfacial region are shown in green, cf. Fig.~\ref{fig3}. AT $\rho =\rho_2$, the shape of the crystalline nucleus changes from spherical to cylindrical, while at $\rho = \rho_3$ a slab configuration (with two planar interfaces) sets in. In the density region where the slab configuration prevails, $\mu = \mu_{coex}$ (as well as $p = p_{coex}$) hold also for finite volumes.}
\end{figure}

When we try to do this at significantly smaller $\eta$ (but still in the two-phase coexistence region, e.g. for $\eta = 0.520$), we regularly find that the crystalline nucleus melts, and the system ultimately becomes a homogeneous fluid; when we try to do this at significantly larger $\eta$, e.g. for $\eta = 0.535$, we find that the compact droplet undergoes ultimately always a fluctuation to an elongated, more or less cylindrical, shape, connected into itself through the periodic boundary condition. For still larger $\eta$ ($0.54 \leq \eta \leq 0.57$), even the cylinder is unstable, and the system develops towards a ``slab'' configuration, where a crystal domain extending in two directions throughout the box is separated from the liquid by two planar interfaces (in this region one must work with $L\times L \times L_z$ boxes, varying $\eta$ by changing $L_z$ and choosing $L$ commensurate with an integer number of lattice planes, of course, to avoid that the crystal domain is strained, cf. Sec.~2.2). Thus, there is only a rather small range of packing fractions where the desired equilibrium between a crystalline nucleus and surrounding fluid is actually stable.

A qualitative explanation of the situation is provided in Fig.~\ref{fig9}, where the chemical potential is plotted versus density: in a finite system, the region of two-phase coexistence actually is reduced in comparison with the bulk. The homogeneous liquid in fact loses its thermodynamic stability not for $\rho = \rho_f$ but at the higher density $\rho_1$. This enhanced stability of the fluid for densities where in the thermodynamic limit the fluid would be metastable is due to the fact that the surface free energy cost associated with the formation of a nucleus would be too high. The same phenomena is well know for the vapor-liquid transition \cite{53,54,55,56,57,58}, where the transition at $\rho = \rho_1$ is termed the droplet evaporation-condensation transition \cite{53,54,55,80}. It has been suggested that \cite{55,80} $\rho_1 - \rho_f \propto V_{box}^{-1/4}$ in $d=3$ dimensions, while the densities $\rho_2, \rho_3$ show only insignificant dependence on box size.

A plot of pressure $p$ versus density looks similar to Fig.~\ref{fig9} when the plot refers to the majority phase, i.e. the liquid phase for $\rho_f < \rho <\rho_d = (\rho_f + \rho_m)/2$, and the crystal for $\rho_d < \rho < \rho_m$. Note that the interchange of what majority and minority means, at $\rho = \rho_d$, happens in the regime where the slab morphology prevails, and then $p_\ell = p_c =p_{coex}$. On the other hand, in the regions where the fluid coexists with compact nuclei (spherical droplets, in the vapor to liquid transition, respectively) the pressure inside the nucleus is enhanced in comparison with the pressure n the surrounding phase. While in situations of phase coexistence the chemical potential is homogeneous (as it must be in equilibrium, when the coexisting domains may exchange particles freely), the pressure distribution is not homogeneous. Even for the slab configuration, where the pressure inside both coexisting phases is the same, as stated above, a nontrivial variation of the pressure tensor components occurs in the interfacial regions.

\begin{figure}
\centering
\begin{minipage}[t]{0.25\textwidth}
\includegraphics[width=\textwidth]{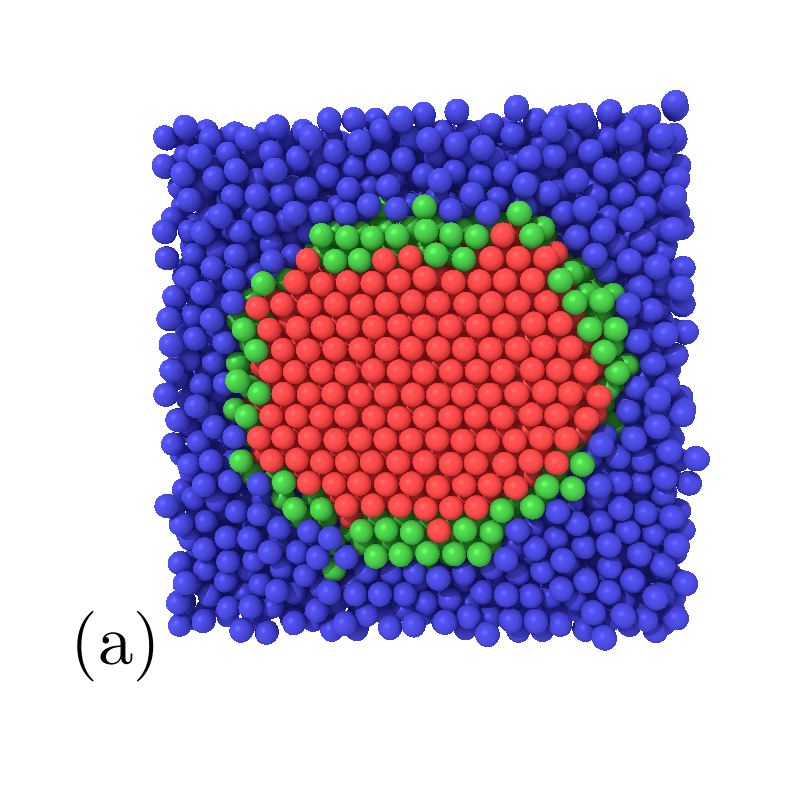}
\end{minipage}%
\begin{minipage}[t]{0.25\textwidth}
\includegraphics[width=\textwidth]{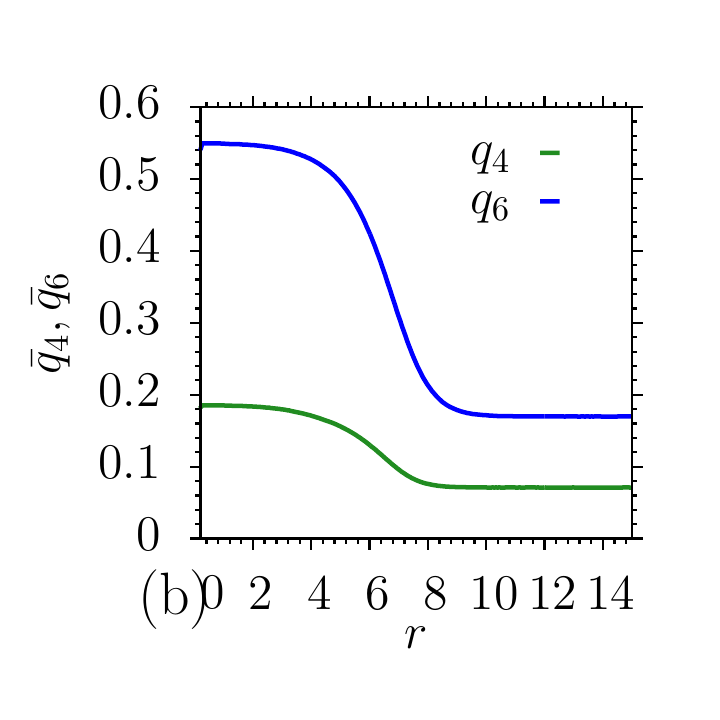}
\end{minipage}
\caption{\label{fig10} (a) Cutout of the cross section through the system, for a plane containing the center of mass of the (compact) crystalline nucleus for the choice $\eta = 0.527$ of the total packing fraction and $N=8000$. Particles belonging to the crystal are shown in red; their regular arrangement can be clearly recognized. Particles clearly belonging to the fluid are shown in blue, while green particles belong to the interfacial region. The classification of solid and liquid can be found in \cite{color}.
(b) Radial density profiles of the order parameters $\tilde{q}_4$ and $\tilde{q}_6$, for the case where $N=8000, \eta = 0.528$ and the packing fraction in the liquid could be estimated as $\eta_l = 0.512$ at a pressure of $p_l=9.74$.}
\end{figure}

In this region of packing fractions $\eta$, where a stable equilibrium between a compact nucleus and surrounding fluid is established we record after every $10\cdot10^6$ Monte Carlo steps the configuration of the crystalline nucleus, see Fig.~\ref{fig10}a for an example, and take an average over the radial density distribution of the particles belonging to this nucleus around its center of mass (Fig.~\ref{fig10}b). Of course, we do not imply that the shape of the crystalline nucleus is on average spherical (although this turns out to be a rather good approximation). But the information shown in Fig.~\ref{fig10}a,b helps us in each configuration of the system that is analyzed, to identify the region of the fluid that is sufficiently
remote from the crystal (where $q_4,q_6$ in Fig.~\ref{fig10}b are already flat, independent of $r$, and have values typical for the liquid phase), where now measurements of the pressure $p_\ell$ and associated packing fraction $\eta_\ell (p_\ell)$ can be performed. Gratifyingly, these data (Fig.~\ref{fig11}) are fully compatible with the continuation of the liquid branch of the isotherm (Fig.~\ref{fig2}) for pressures exceeding $p_{coex}$. This result shows that the liquid surrounding the nucleus is in thermal equilibrium, which is a basic condition for the analysis presented in the following section.

\section{Estimation of the surface excess free energy of crystalline nuclei surrounded by liquid from a thermodynamic analysis}

\begin{figure}
\centering
\includegraphics[width=0.47\textwidth]{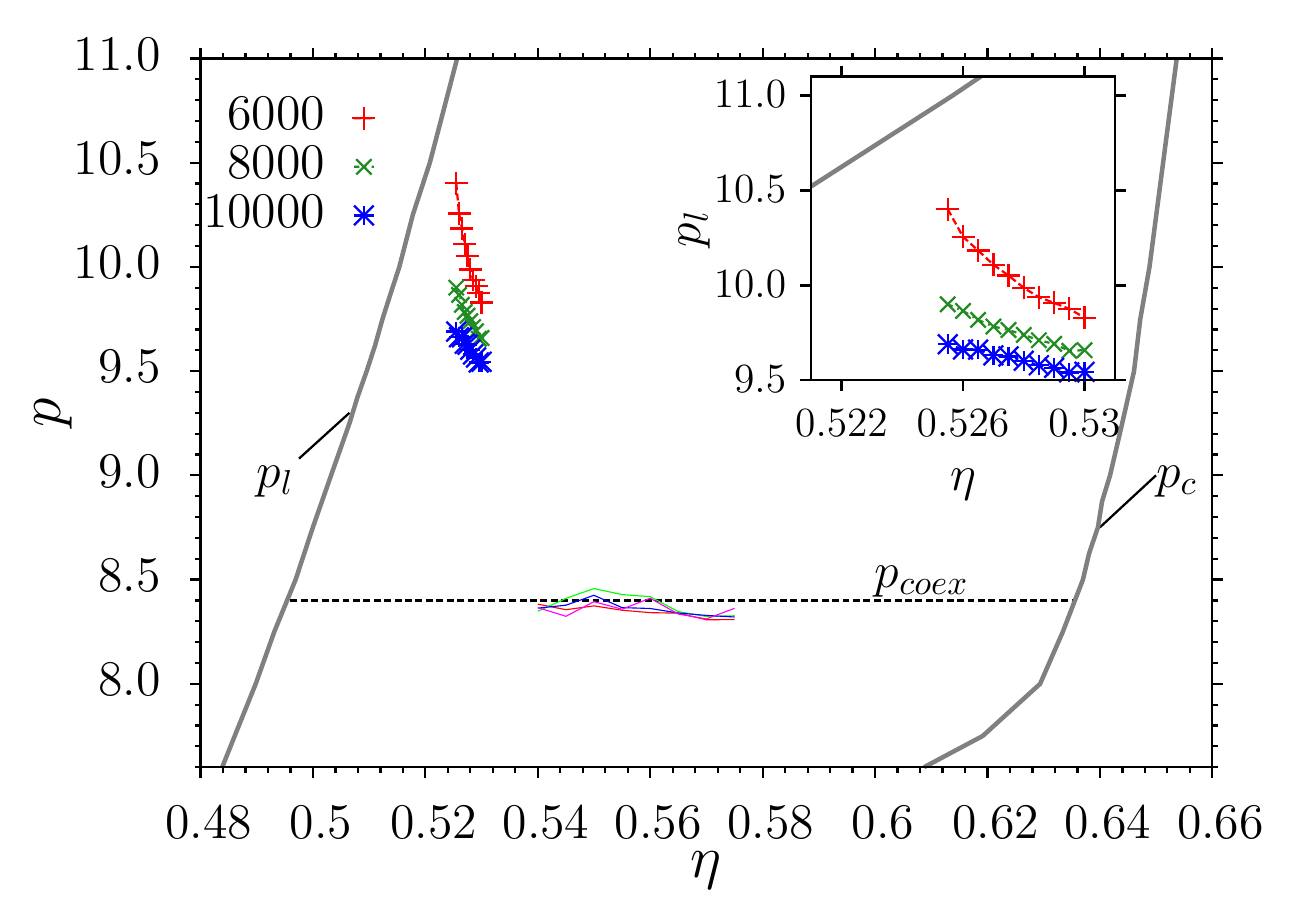}
\caption{\label{fig11} Pressure $p_\ell$ of the liquid surrounding a droplet plotted vs. average packing fraction $\eta$ of the colloid particles in the total simulation volume. Three choices of particle number $N$ are included, as indicated. For clarity, the region of interest is shown also on a plot with strongly magnified abscissa scale. Also data for $0.54 < \eta < 0.575$ are included, where the crystal forms a slab configuration (cf. Fig.~9), and the pressure agrees (apart from fluctuations) with $p_{coex}$ as it should be.}
\end{figure}

\begin{figure}
\centering
\includegraphics[width=0.35\textwidth]{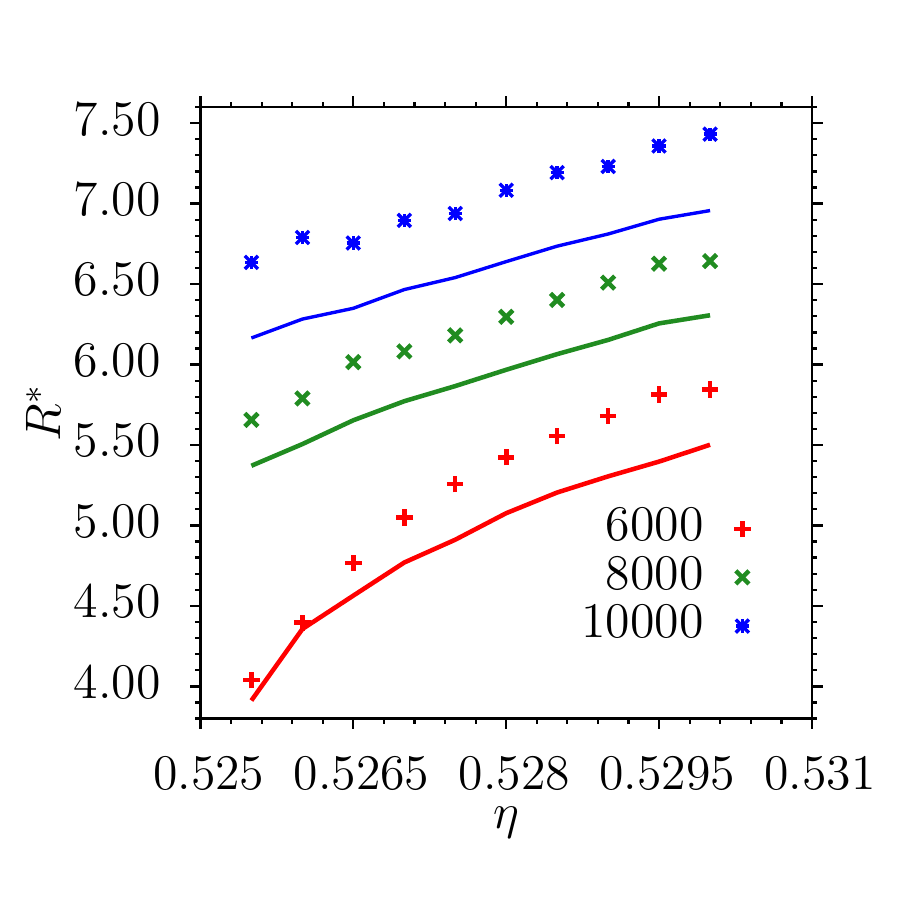}
\caption{\label{fig12} (a) Nucleus radius $R^*$ extracted from the inflection point of the order parameter profile $q_6(r)$ [within error it coincides with inflection point of $q_4(r)$] plotted vs. average packing fraction.}
\end{figure}

Fig.~\ref{fig11} now shows our data for the pressure of the fluid phase surrounding the nucleus for the three choices of particle number N = 6000, 8000 and 10 000, in the regime $0.525  < \eta < 0.530$, where stable nuclei occur. We now use a simple extension of the lever rule for phase coexistence, adapted to account for finite size effects, to infer the volume $V^*$ of the crystalline nucleus, namely
\begin{equation}\label{eq17}
\eta V_{box} = \eta_\ell (p_\ell) (V_{box} - V^*) + \eta_c (p_c) V^*
\end{equation}
Note that $\eta$ is the chosen control variable and when $N$ is chosen also $V_{box}$ is fixed. We observe $p_\ell$ as well as $\eta_\ell(p_\ell)$ directly, as discussed above. Since a direct estimation of $\eta_c(p_c)$ in a small fluctuating crystal nucleus suffers from relatively large statistical errors (and perhaps also systematic errors, if parts of the interfacial region are included), we use the appropriate value $\eta_c(p_c)$ from the bulk crystal branch of the isotherm (Fig.~\ref{fig2}). To find $\eta_c(p_c)$ we use the fact that we know $\mu_{coex}$ to carry out the thermodynamic integration.
\begin{equation}\label{eq18}
\mu_c (p_c) = \mu_{coex} + \frac \pi 6 \int \limits _{p_{coex}}^{p_c} dp / \eta_c(p)
\end{equation}
to find the chemical potential of the crystal branch of the isotherm. From Figs.~\ref{fig2},\ref{fig11} we note that $\eta_c(p)$ is a slightly nonlinear function of $p$ near $p_{coex}$, unlike $\eta_\ell(p)$, and therefore it is necessary to go beyond the simple linear expansion (cf.~Eq.\ref{eq13}) $\mu_c(p_c) \approx \mu_{coex} + \frac \pi 6 \frac {1}{\eta_m}(p_c - p_{coex}$) (note $\eta_m = \eta_c(p_{coex})$.) The crystal pressure $p_c $ then follows from Eq.~\ref{eq8} since we have determined $\mu$ and $p_\ell$ in the liquid region explicitly, the equality of the chemical potentials determines also $p_c$ through Eq.~\ref{eq18}. Using then the data of Fig.~\ref{fig2}, where the pressure of the crystal $p_c$ is determined as function of its packing fraction, we obtain $\eta_c(p_c)$ once $p_c$ has been obtained. Then the crystallite volume $V^*$ is obtained from Eq.~\ref{eq17}, without the need of assuming anything about the shape of this crystal nucleus. If we would assume that the volume is strictly spherical, $V^*=4\pi R^{*3}/3$, we could try to use the radial order parameter profiles of $q_4(r)$ and $q_6(r)$, Fig.~\ref{fig10}b, to estimate $R^*$ from the inflection point of these curves. Fig.~\ref{fig12} shows effective radii obtained in this way, and compares them with an effective radius $R^*_{eff}= (3V^*/4\pi)^{1/3}$ extracted from the general method (using Eqs.~\ref{eq17},\ref{eq18}) with the spherical approximation. Unexpectedly, these data indicate a rather good agreement, indicating that for the present model anisotropy effects of the interface tension cannot be very important.

To elaborate the consequences of this finding for the theory of homogeneous nucleation of crystals in this system, we note that in classical nucleation theory the formation free energy of a nucleus is written in terms of volume and surface terms as \cite{1,2,3,4,5,6,7}
\begin{equation}\label{eq19}
\Delta F(V^*)= - (p_c-p_\ell)V^* + F_{surf}(V^*)
\end{equation}

In the description of the surface excess free energy $F_{surf}(V^*)$ of the crystal nucleus, the dependence of the crystal-liquid interface tension $\gamma(\vec{n})$ on the orientation of the interface normal $\vec{n}$ relative to the crystal axes \cite{29,30,31,32,33,34,35,36} needs to be taken into account. Only for isotropic $\gamma$ the nucleus is a sphere of radius $R$ and this surface excess free energy simply is $F_{surf} = 4 \pi R^2\gamma = A_{iso} \gamma V^{2/3}$ with a geometric factor $A_{iso}=(36 \pi)^{1/3}$. For crystals, however, the term $A_{iso}\; \gamma$ has to be replaced by a more complicated expression
\begin{equation}\label{eq20}
F_ {surf} (V) = V^{2/3} \int \limits _{A_w}  \gamma (\vec{n}) d\vec{s} \equiv A_w \bar{\gamma} V^{2/3}\;,
\end{equation}
where $A_w$ is the surface area of a unit volume whose shape is derivable from $\gamma (\vec{n})$ with the Wulff construction \cite{30,31,32,33,34,35,36}, and the integration in Eq.~\ref{eq20} is extended over this surface. The average interface tension $\bar{\gamma}$ introduced in Eq.~\ref{eq20} is defined as
\begin{equation}\label{eq21}
\bar{\gamma} = A_w^{-1} \int \gamma (\vec{n}) ds\quad .
\end{equation}
The critical nucleus according to the theory of homogeneous nucleation satisfies the conditions $\partial(\Delta F(V^*))/\partial V^*=0$, and hence Eqs.~\ref{eq19}, \ref{eq20} yield
\begin{equation}\label{eq22}
V^* = [\frac{2A_w\bar{\gamma}}{3(p_c-p_\ell)}]^{1/3}
\end{equation}
and the associated free energy barrier is
\begin{equation}\label{eq23}
\Delta F^* = \frac 1 3 A_w \bar{\gamma} V^{*2/3} = \frac 1 2 (p_c-p_\ell)V^*\quad .
\end{equation}

While in the thermodynamic limit the liquid surrounding the critical nucleus is metastable, and actually the equilibrium condition $\partial (\Delta F(V^*)/\partial V^*)=0$ refers to the unstable equilibrium on top of a saddle point in configuration space, we here deal with a stable equilibrium between the nucleus and the surrounding nucleus in the finite box. The equilibrium condition is the same, however, because in both cases the nucleus can freely exchange particles with the surrounding liquid, and the chemical potential in the system is constant. Of course, the analysis presented cannot be extended to arbitrary small box volumes, since working at fixed packing fraction $\eta = (\pi/6) N \sigma_c^3 /V_{box}$ the chemical potential in a small volume is not strictly constant, but exhibits statistical fluctuations which are not independent from the statistical fluctuations of $V^*$, and hence give rise to finite size effects.

However, an attractive feature of Eq.~\ref{eq23} is that actually neither a knowledge of $A_w$ nor of $\bar{\gamma}$ is needed to predict $\Delta F^*$: knowing $V^*$ and the pressure difference $p_c - p_\ell$ yields directly $\Delta F^*$, and hence the need to estimate $\gamma(\vec{n})$ and carry out the Wulff construction is bypassed.

\begin{figure}
\centering
\includegraphics[width=0.3\textwidth]{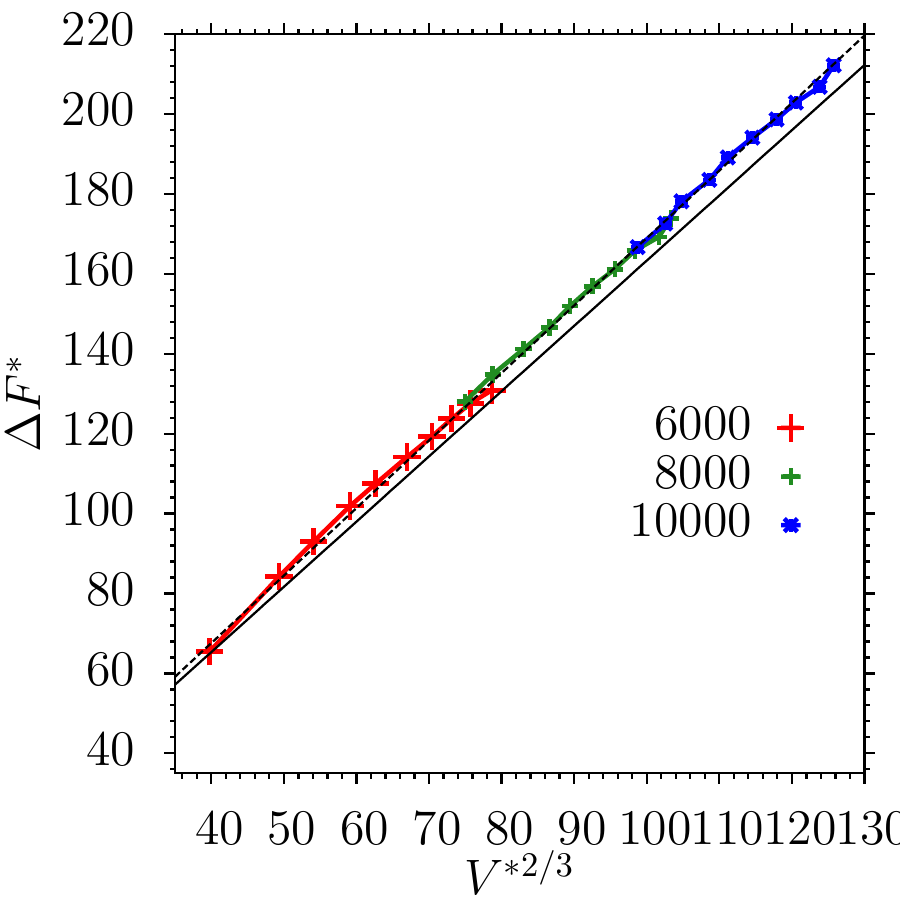}
\caption{\label{fig13} Barrier $\Delta F^*$ against homogeneous nucleation (in units of $k_BT$) plotted vs. $V^{*2/3}$, using the data for $V^*$ and $p_c - p_\ell$ from the thermodynamic analysis of the equilibrium between nucleus and surrounding liquid, for the data for 0.525 $\leq \eta \leq 0.530$ and $N = 6000, 8000 $ and $ 10 000$. The straight line is a fit through the data that includes the origin $(\Delta F^* = 0 $ for $V^* =0)$, using an independent estimate for the surface tension and assuming a spherical nucleus shape  (see text).}
\end{figure}

Fig.~\ref{fig13} shows a plot of $\Delta F^*$ vs. $V^{*2/3}$, using our simulation results for the three different choices of $N$. Gratifyingly, the data for different $N$ overlap, there do not seem to be significant finite size effects present (at least not for the regime of $V^*\geq 250$ explored here). It also is interesting that the data do not indicate significant deviations from the straight line behavior, indicating that we get a meaningful estimate of the constant $A_w \bar{\gamma}/3$ from the slope of this straight line. Unlike related work on vapor to liquid nucleation in simple fluids \cite{56,57,58}, curvature corrections (of order $V^{*1/3}$ in this plot) do not seem to play any role here, thus there is also no need to consider the distinction between the equimolar dividing surface (implicit in Eq.~\ref{eq17}) and the surface of tension which needed to be considered in the vapor to liquid case \cite{57}).

\begin{figure}
\centering
\includegraphics[width=0.3\textwidth]{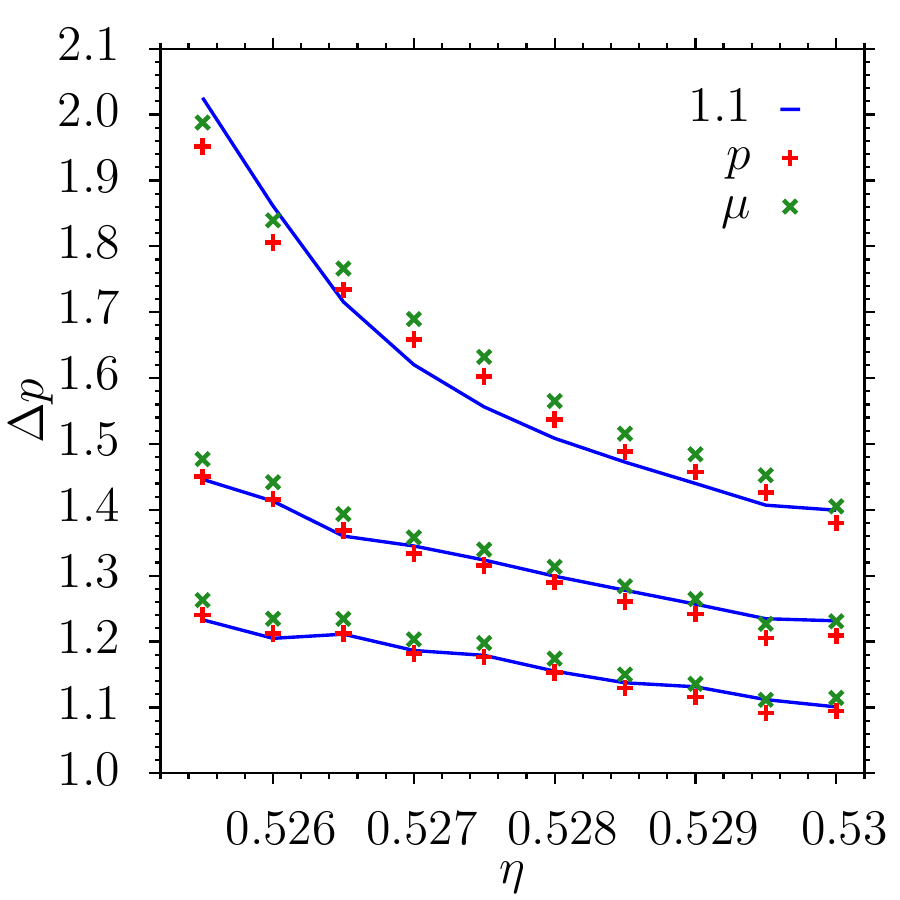}
\caption{\label{fig14}
Pressure difference $ p_\ell - p_{coex} $ plotted versus the average packing fraction $\eta$ in the simulation box, for particle number N = 6000 (topmost set of data), 8000 (middle set of data) and 10 000(lowest set), respectively. Data points are from the thermodynamic analysis (green symbols represent the formula $ p_\ell - p_{coex} = \frac 6 \pi \eta _f(\mu_f-\mu_{coex})$), curves show the prediction $p_\ell - p_{coex} = c \cdot (2 \gamma_{111}/R)/(\eta_m/\eta_f-1)$, using $\gamma_{111} \approx 1.013$ (taken from \cite{81}) and $c=1.07$ as indicated.}
\end{figure}

Since we have already seen that the direct estimation of nuclei volume from a spherical approximation (Fig.~\ref{fig12}) is in good agreement with the volume determined from the thermodynamic analysis, it is of course tempting to try a direct spherical approximation of $\Delta F^*$ as well, replacing $A_w$ in Eq.~\ref{eq21} by $A_{iso}=(36 \pi)^{1/3}$, and using instead of $\bar{\gamma}$ an estimate for the smallest interface tension \cite{81}, which occurs for the close-packed 111 crystal surface, namely $\gamma_{111} = 1.013$(3), yielding $A_{iso} \gamma/3 \approx 1.633.$ It turns out that the resulting straight line is almost indistinguishable from the straight line included in Fig.~\ref{fig13}! This good agreement can also be seen when one considers the pressure difference $\Delta p = p_c-p_{coex}$ as function of $\eta$ (Fig.~\ref{fig14}) and includes there the prediction appropriate for a spherical nucleus, $\Delta p = (2 \gamma/R^*)/(\frac{\eta_m}{\eta_f} - 1)$ for the individual choices of $N$, using again $\gamma_{111}$ as an approximation for $\gamma$. Although the other interface tensions $\gamma_{100}, \gamma_{110}$ for the present model are about 3\% larger than $\gamma_{111}$ \cite{81}, this weak anisotropy of the present model does not lead to significant deviations from the spherical approximation, for the present choice of model parameters.

\section{Conclusions}

In this paper we have introduced a smooth variant of the Asakura-Oosawa model for colloid-polymer mixtures, in the limit of small size ratios where the polymers can be integrated out in favor of a short range colloid-colloid attraction, to study the phase transition from the fluid phase to the face-centered cubic crystal. We have paid particular attention to phase coexistence in finite volumes, and the information on nucleation barriers that one can extract from an analysis of phase coexistence. A key ingredient of the discussion has been the observation that the chemical potential of the system, considered as a function of packing fraction at constant volume, exhibits a loop (Fig.~\ref{fig9}). The enhancement of $\mu$ relative to $\mu_{coex}$ in the first descending part of this loop, and the associated enhancement of the pressure $p_\ell$ in the liquid over $p_{coex}$, contains information that can be used to estimate the surface excess free energy of the crystalline nucleus.

In order to be able to study this problem with meaningful accuracy, we have first located phase coexistence between bulk phases from the standard ``interface velocity'' method, where one prepares a ``slab configuration'' of a crystal coexisting with liquid, separated from it by planar interfaces, in a simulation box of suitably elongated shape. Studying the ``velocity'' with which the volume of the box changes with time (due to interface motion, when the crystal shrinks for $p < p_{coex}$ or grows for $p>p_{coex}$) , one can estimate the coexistence pressure $p_{coex}$. Unfortunately, this method due to large statistical fluctuations (Fig.~\ref{fig4}) and finite size effects (Fig.~\ref{fig5}) requires a major computational effort, and alternative more efficient methods to accurately estimate $p_{coex}$ would be desirable. After this part of our work was completed, a method due to Pedersen et al. \cite{87} came to our attention, which could be advantageous for this purpose.

We have also applied a method where the fluid in an elongated simulation box is exposed to one wall with a very soft potential, enabling the estimation of the chemical potential $\mu$ also in a situation where in the bulk the packing fraction is too high so that the standard Widom particle insertion method would not work (Fig.~\ref{fig6},\ref{fig7}). We have shown that by this method $\mu$ can be estimated for the liquid branch of the isotherm, even for pressures $p>p_{coex}$.

The crystalline nuclei (coexisting with fluid) have been studied both by geometrical identification where the crystal-fluid interface is located (Fig.~\ref{fig10},\ref{fig12}) using the well-known concept of bond orientational order parameters $(\tilde{q}_4,\tilde{q}_6$, cf. Fig.~\ref{fig8}) to locally distinguish the phases, and by thermodynamic analysis (Figs.\ref{fig11},\ref{fig13},\ref{fig14}). We show that for the present model the assumption that the anisotropy of $\gamma(\vec{n})$ can essentially be ignored, and therefore the nuclei have effectively a spherical shape, works very accurately. We stress, however, that we do not expect that this conclusion works in general; if the anisotropy $\gamma(\vec{n})$ is more pronounced, the spherical approximation must fail, but our thermodynamic method which only needs the information on the fluid pressure, fluid chemical potential, and nucleus volume, still should work.

Such a more anisotropic case is expected for the present model at large $z_p$ in Eq.~\ref{eq1}b, when $\eta_f$ is much smaller than $\eta_m$. However, an explicit study of this case must be left to future work.

An important feature of our work also is the excellent agreement with classical nucleation theory, $\Delta F^* = A_w \bar{\gamma} V^{*2/3}/3$ (Fig.~\ref{fig13}). This conclusion seems to be at variance with both experimental and simulation studies on nucleation in colloidal suspensions (see \cite{19} for a review). However, we feel that most of this work has been done for conditions where $(\eta_\ell - \eta_f)/\eta_f$ is much larger than in the present work, so actually the nucleation barrier is then rather small; our pressure range $9.54 \leq p\sigma^3_c /k_BT \leq 10.40$ for the pressure of the metastable fluid (Fig.~\ref{fig11}) translates into packing fractions $0.51 \leq \eta \leq 0.52$  of the fluid surrounding the crystalline nucleus in our box. These numbers are close to $\eta_f = 0.494$, the relative distance from the transition where freezing sets in is at most 6\%. In contrast, experiments \cite{9} and simulations for hard spheres \cite{82,83,84} typically deal with packing fractions in the range $0.53 < \eta < 0.57$ (note that within the available accuracy $\eta_f$ in our model agrees with the corresponding value for hard spheres). Note that this range of the experiments corresponds to huge values of $p_\ell - p_{coex}$ (cf.~Fig.~\ref{fig2}), and correspondingly the estimate

\begin{equation}\label{eq24}
R^* = 2 \gamma/[(p_\ell - p_{coex})(\eta_m/\eta_f-1)]
\end{equation}

that classical nucleation theory would imply yields droplet radii that do not exceed about two colloid diameters, and barriers that are only of the order of 5 to 10 $k_BT$. It is clear that for such small nuclei the approximations of the classical nucleation theory are expected to fail. On the other hand, for such dense colloidal suspensions the nucleation rate will be strongly affected by the slowing down that sets in when the colloidal glass transition \cite{85,86} is approached. In fact, Radu and Schilling \cite{27} have pointed out a strong and nontrivial effect due to the kinetic prefactor of the nucleation rate. Since the slowing down associated with the glass transition is not a well understood problem \cite{42}, studies attempting to test nucleation theory should avoid this region, and focus on the region close to $\eta_f$, as done here.
Finally, we discuss the issue why the spherical approximation is such a good approximation to the actual crystal shape in this case.
The simple explanation is that the anisotropy pf the interfacial tension of the crystal fluid interface in this model is very weak, the difference
between $\gamma_{111}, \gamma_{110}$ and $\gamma_{100}$ is a few \% only. A similar situation also arises in the lattice gas model \cite{28}, at temperatures well above the roughening temperature. Recent work \cite{88} has studied the angular dependence of the interface tension in this model over a wide range of temperatures, showing that the dependence on interface tension on interface orientation amounts to a variation of typically a few \% as well (this anisotropy vanishes at the critical point of this model). The corresponding enhancement of the nucleation barrier was found to be of the order of 10 \% as well \cite{28}; only at low temperatures, where facetted crystal shapes occur, large derivations from the spherical approximations were found \cite{28}. Thus, we expect that the present findings are generic for all systems where the anisotropy of the interfacial tension at melting conditions amounts only a few \%.

\underline{Acknowledgements}: This work was supported by the Deutsche Forschungsgemeinschaft (DFG) under grant No's SFB-TR6/D5 and VI237/4-3. We thank the Hochleistungsrechenzentrum Stuttgart (HLRS) and the Zentrum f\"ur Datenverarbeitung Mainz for generous grants of computing time at the HERMIT and MOGON supercomputers.


\begin{thebibliography}{99}
\bibitem{1} A. C. Zettlemoyer (ed.) \textit{Nucleation} (M. Dekker, New York, 1969)
\bibitem{2} W. A. Tiller, \textit{The Science of Crystallization: Microscopic Interfacial Phenomena} (Cambridge Univ. Press, New York, 1991)
\bibitem{3} I. Gutzow and J. Schmelzer, \textit{The Vitreous State: Thermodynamics, Structure, Rheology, and Crystallization} (Springer, Berlin, 1995)
\bibitem{4} D. Kashchiev, \textit{Nucleation: Basic Theory with Applications} (Butterworth-Heinemann, Oxford, 2000)
\bibitem{5} S. Balibar and F. Caupin, Comptes Rendus Physique \textbf{7}, 988 (2006)
\bibitem{6} D. M. Herlach, P. Galenko, and D. Holland-Moritz, \textit{Metastable Solids from Undercooled Melts} (Pergamon, Oxford, 2007)
\bibitem{7} K. F. Kelton and A. I. Greer, \textit{Nucleation} (Pergamon, Oxford, 2009)
\bibitem{8} H. Emmerich, P. Virnau, G. Wilde and R. Spatschek (Eds.) \textit{Heterogeneous Nucleation and Microstructure Formation: Steps Towards a System and Scale Bridging Understanding}, EPJ \textbf{223}, No3 (2014)
\bibitem{9} T. Palberg, J. Phys.: Condens. Matter \textbf{26}, 333101 (2014)
\bibitem{10} H. M\"uller-Krumbhaar, W. Kurz, and E. Brener, in \textit{Phase Transformations in Materials} (G. Kostorz, ed.) p. 81 (Wiley-VCH, Weinheim, 2001)
\bibitem{11} D. Turnbull, J. Appl. Phys. \textbf{21}, 1022 (1950)
\bibitem{12} J. E. Burke and D. Turnbull, Progr. Met. Phys. \textbf{3} 220 (1952)
\bibitem{13} H. Biloni, in \textit{Physical Metallurgy} (R. W. Cahn and P. Haasen, eds.) p. 477 (North-Holland, Amsterdam, 1983)
\bibitem{14} W. J. Boettinger, S. R. Correll, A. L. Greer, A. Karma, W. Kurz, M. Rappaz, and R. Trivedi, Acta Metr. \textbf{48}, 43 (2000)
\bibitem{15} S. Auer and D. Frenkel, Condens. Matter \textbf{14}, 7667 (2002)
\bibitem{16} A. Cacciuto, S. Auer and D. Frenkel, Phys. Rev. Lett. \textbf{93}, 166105 (2004)
\bibitem{17} A. Cacciuto, S. Auer and D. Frenkel, J. Chem. Phys. \textbf{119}, 7467 (2003)
\bibitem{18} D. Winter, P. Virnau and K. Binder, Phys. Rev. Lett. \textbf{103}, 225703 (2009)
\bibitem{19} D. Winter, P. Virnau and K. Binder, J. Phys.: Condens. Matter \textbf{21}, 464118 (2009)
\bibitem{20} D. Deb, A. Winkler, P. Virnau, and K. Binder, J. Chem. Phys. \textbf{136}, 134710 (2012)
\bibitem{21} S. Dorosz, T. Schilling, J. Chem. Phys. \textbf{136}, 044702 (2012)
\bibitem{22} V. W. A de Villeneuve, D. Verboekend, R. P. A. Dullens, D. G. A. L. Aarts, W. K. Kegel, and H. N. W. Lekkerkerker, J. Phys.: Condens. Matter \textbf{17}, S3371 (2005)
\bibitem{23} H. J. Sch\"ope and P. Wette, Phys. Rev. E\textbf{83}, 051405 (2011)
\bibitem{24} S. Jungblut and C. Dellago, EPL \textbf{96}, 56006 (2011)
\bibitem{25} A. Engelbrecht and H. J. Sch\"ope, Soft Matter \textbf{8}, 11034 (2012)
\bibitem{26} K. Binder and D. Stauffer, Adv. Phys. \textbf{25}, 343 (1976)
\bibitem{27} M. Radu and T. Schilling, EPL \textbf{105}, 26001 (2014)
\bibitem{28} F. Schmitz, P. Virnau, and K. Binder, Phys. Rev. E\textbf{81} 705330 (2013)
\bibitem{29} G. Wulff, Z. Kristallogr. Mineral. \textbf{34}, 449 (1901)
\bibitem{30} C. Herring, Phys. Rev. \textbf{82}, 87 (1951)
\bibitem{31} C. Herring, in \textit{Structure and Properties of Solid Surfaces} (R. Gomer, ed.) p. 5 (Univ. of Chicago Press, Chicago, 1952)
\bibitem{32} J. W. Cahn and D. W. Hoffmann, Acta Metall. \textbf{22}, 1205 (1974)
\bibitem{33} D. W. Hoffmann and J. W. Cahn, Surf. Sci \textbf{31}, 368 (1972)
\bibitem{34} R. K. P. Zia and J. Avron, Phys. Rev. B\textbf{25}, 2042 (1982)
\bibitem{35} C. Rottmann and M. Wortis, Phys. Rep. \textbf{103}, 59 (1984)
\bibitem{36} M. Wortis, in \textit{Chemistry and Physics of Solid Surfaces, Vol. II} (R. Vanselow, ed.) (Springer, Berlin, 1988)
\bibitem{37} K. Binder, PNAS \textbf{111}, 3974 (2014)
\bibitem{38} P. N. Pusey and W. van Megen, Nature \textbf{320}, 340 (1986)
\bibitem{39} P. N. Pusey, in \textit{Liquids, Freezing, and the Glass Transition,} edited by J. P. Hansen, D. Levesque, and J. Zinn-Justin (North-Holland, Amsterdam, 1991) p. 736.
\bibitem{40} W. C. K. Poon and P. N. Pusey, in \textit{Observation, Prediction and Simulation of Phase Transitions in Complex Fluids}, edited by M. Baus, L. F. Rull, and J. P. Ryckaert (Kluwer, Dordrecht, 1995) p. 3
\bibitem{41} H. Bach and D. Krause (eds.) \textit{Analyses of the Composition and Structure of Glass and Glass Ceramics} (Springer, Berlin, 1999)
\bibitem{42} K. Binder and W. Kob, \textit{Glassy Materials and Disordered Solids: An Introduction to Their Statistical Mechanics}, 2nd ed. (World Scientific, Singapore, 2011)
\bibitem{43} A. Ha, I. Cohen, X. Zhao, M. Lee, and D. Kivelson, J. Phys. Chem. \textbf{100}, 1 (1996)
\bibitem{44} H. Tanaka, Eur. Phys. J. E\textbf{35}, 113 (2012)
\bibitem{45} R. Boehmer, G. Hinze, G. Diezemann, B. Geil, and H. Sillescu, Europhys. Lett. \textbf{36}, 55 (1996)
\bibitem{46} H. E. Stanley (ed.) \textit{Liquid Polymorphism} (J. Wiley \& Sons, Hoboken, 2014)
\bibitem{47} A. Statt , P.Virnau and K. Binder , Phys. Rev. Lett \textbf{114}, 026101 (2015)
\bibitem{48} S. Asakura and F. Oosawa, J. Chem. Phys. \textbf{22}, 1255 (1954)
\bibitem{49} S. Asakura and F. Oosawa, J. Polym. Sci. \textbf{33}, 183 (1998)
\bibitem{50} K. Binder, P. Virnau, and A. Statt,  J. Chem. Phys. \textbf{14}, 141 (2014)
\bibitem{51} W. C. K. Poon, J. Phys.: Condens. Matter \textbf{14}, R859 (2002)
\bibitem{52} H. N. W. Lekkerkerker and R. Tuinier, \textit{Colloids and the Depletion Interaction} (Springer, Berlin, 2011)
\bibitem{53} L. G. McDowell, P. Virnau, M. M\"uller, and K. Binder, J. Chem. Phys. \textbf{120}, 5293 (2004)
\bibitem{54} L. G. McDowell, V. K. Shen, and J. R. Errington, J. Chem. Phys. \textbf{125}, 034705 (2006)
\bibitem{55} M. Schrader, P. Virnau and K. Binder, Phys. Rev. E\textbf{79}, 061104 (2009)
\bibitem{56} B. J. Block, S. K. Das, M. Oettel, P. Virnau, and K. Binder, J. Chem. Phys. \textbf{135}, 154702 (2010)
\bibitem{57}  A. Troester, M. Oettel, B. J. Block, P. Virnau and K .Binder, J. Chem. Phys. \textbf{136}, 064709 (2012)
\bibitem{58} K. Binder, B. J. Block, P. Virnau and A. Troester, Am. J. Phys. \textbf{80}, 1099 (2012)
\bibitem{59} P. J. Steinhardt, D. R. Nelson, and M. Ronchetti, Phys. Rev. B\textbf{28}, 783 (1983)
\bibitem{60} W. Lechner and C. Dellago, J. Chem. Phys. \textbf{129}, 114707 (2008)
\bibitem{61} J. D. Weeks, J. Chem. Phys. \textbf{67}, 3106 (1977)
\bibitem{62} J. S: Rowlinson and B. Widom, \textit{Molecular Theory of Capillarity} (Clarendon Press, Oxford, 1982)
\bibitem{63} K. Binder, M. M\"uller, F. Schmid and A. Werner, Adv. Colloid Interface Sci., \textbf{94}, 237 (2001)
\bibitem{64} P. Tarazona and E. Chacon, Phys. Rev. B\textbf{70}, 235407 (2004)
\bibitem{65} T. Zykova-Timan, J. Horbach, and K. Binder, J. Chem. Phys. \textbf{133}, 014705 (2010)
\bibitem{66} K. Binder, J. Phys.: Conf. Series \textbf{510}, 012002 (2014)
\bibitem{67} P. J. Flory, \textit{Principles of Poylmer Chemistry} (Cornell University Press, Ithaca, N. Y., 1953)
\bibitem{68} M. Dijkstra, R. van Roij, and R. Evans, Phys. Rev. E\textbf{59}, 5744 (1999)
\bibitem{69} R. Evans, in \textit{Fundamentals of Inhomogeneous Fluids}, edited by D. Henderson (Dekker, New York, 1992), p. 85
\bibitem{70} J. P. Hansen and I. R. McDonald, \textit{Theory of Simple Liquids} (Academic Press, San Diego, 1980)
\bibitem{71} E. De Miguel and G. Jackson, Mol. Phys. \textbf{104}, 3717 (2006)
\bibitem{72} D. Deb, D. Wilms, A. Winkler, P. Virnau, and K. Binder, Int. J. Mod. Phys. C\textbf{23}, 1240011 (2012)
\bibitem{73}  M. P. Allen and D. J. Tildesley, \textit{Computer Simulation of Liquids} (Clarendon Press, Oxford, 1989)
\bibitem{74} D. C. Rapaport, \textit{The Art of Molecular Dynamics Simulation}, 2nd ed. (Cambridge Univ. Press, Cambridge, 2004)
\bibitem{75} W. C. K. Poon, E. R. Weeks, and C. P. Royall, Soft Matter \textbf{8}, 21 (2012)
\bibitem{76} C. P. Royall, W. C. K. Poon, and E. R. Weeks, Soft Matter \textbf{9}, 17 (2013)
\bibitem{77} A. Statt, \textit{Dissertation} (Johannes-Gutenberg-Universit\"at Mainz, 2015, unpublished)
\bibitem{78} B. Widom, J. Chem. Phys. \textbf{39}, 2808 (1963)
\bibitem{79} J. G. Powles, B. Holtz, and W. A. B. Evans, J. Chem. Phys. \textbf{101}, 7804 (1994)
\bibitem{80} K. Binder, Physica A\textbf{319}, 99 (2003)
\bibitem{81} F. Schmitz and P.Virnau (submitted)
\bibitem{color}
\begin{minipage}[t]{0.4\textwidth}
{\small
\setlength{\abovedisplayskip}{2pt}
\setlength{\belowdisplayskip}{0pt}
\setlength{\abovedisplayshortskip}{0pt}
\setlength{\belowdisplayshortskip}{0pt}
For distinguishing different phases, the following rules are applied
\begin{align*}
&\text{solid fcc: } \bar q_6 < \pm \sqrt{b_0-(a_0-\bar q_4)^2}+c_0+e_0 \bar q_4 &\\
&\text{with }\; a_0=0.165, \; b_0=0.0011,  c_0=-0.017  e_0=3.00 \; ,&\\
&\text{liquid: } \bar q_6 < \pm \sqrt{b_1-(a_1-\bar q_4)^2}+c_1 &\\
&\text{with }\;    b_1=0.015, \; c_1=0.1800,  a_1=0.08 , \; \text{ and }&\\
&\text{solid hcp: }\bar q_6 >a_2 \bar q_4+b_2 &\\
&\text{with }\;   a_2=2.000, ; b_2=0.2400 \;.&
\end{align*}
with empirically chosen constants. The interface is everything which is not identified as solid or liquid.}
\end{minipage}
\bibitem{82} S. Auer and D. Frenkel, Nature \textbf{409}, 1020 (2001)
\bibitem{83} L. Filion, M. Hermes, R. Ni, and M. Dijkstra, J. Chem. Phys. \textbf{133}, 244115 (2010)
\bibitem{84} T. Schilling, S. Dorosz, H. J. Sch\"ope, and G. Opletal, J. Phys.: Condens. Matter \textbf{23}, 194120 (2011)
\bibitem{85} P. N. Pusey and W. van Megen, Nature \textbf{320}, 340 (1980)
\bibitem{86} P. N. Pusey, in \textit{Liquids, Freezing and Glass Transition}, edited by J. P. Hansen, D. Levesque and J. Zinn-Justin, (Elsevier, Amsterdam, 1991) Chapter 10.
\bibitem{87} U. R. Pedersetn, F. Hummel, G. Kresse, G. Kahl, and C. Dellago, Phys. Rev. B\textbf{88}, 094101 (2013); U. R. Pedersen, J. Chem. Phys. \textbf{139}, 104102 (2013)
\bibitem{88} B. J. Block, S. Kim, P.Virnau, and K. Binder, Phys.Rev E \textbf{90}, 062106 (2014)
\bibitem{fcc1}  L. V. Woodcock, Nature \textbf{385.6612}, 141-143 (1997)
\bibitem{fcc2} Bolhuis, Peter G., D. M. S. C. Frenkel, Siun-Choun Mau, and David A. Huse, Nature  \textbf{388.6639}, 235-236 (1997)
\bibitem{fcc3}  A. D. Bruce, N. B. Wilding, and G. J. Ackland. Phys. Rev. Lett \textbf{79}, 3002 (1997)
\bibitem{fcc4}  A. D. Bruce, Bruce, A. D., A. N. Jackson, G. J. Ackland, and N. B. Wilding, Phys. Rev. E \textbf{61}, 906 (2000)
\end{thebibliography}
\end{document}